\documentclass[12pt]{iopart}

\usepackage{graphicx}
\usepackage{bm}
\usepackage{epstopdf}
\usepackage{amsfonts}
\usepackage{amssymb}
\usepackage{latexsym,enumerate}
\usepackage{epsfig}
\usepackage{ulem}
\usepackage{color}
\usepackage[colorlinks=true,linkcolor=blue,urlcolor=blue,citecolor=blue]{hyperref}
\usepackage{subfigure}

\usepackage{fancyhdr}

\newcommand{\ket}[1]{\left\vert#1\right\rangle}
\newcommand{\bra}[1]{\left\langle#1\right\vert}

\def\beq{\begin{equation}}
\def\eeq{\end{equation}}
\def\beal{\begin{eqnarray}}
\def\ealg{\end{eqnarray}}
\def\bearr{\begin{equation}\begin{array}{ll}}
\def\enarr{\end{array}\end{equation}}

\begin{document}

\title{Non-Equilibrium Thermodynamics of Harmonically Trapped Bosons}
\author{Miguel \'Angel Garc\'ia-March} 
\address{ICFO Institut de Ci\`encies Fot\`oniques, The Barcelona Institute of Science and Technology, Av. C.F. Gauss, 3, E-08860 Castelldefels, Spain}
\author{Thom\'as Fogarty}
\address{Theoretische Physik, Universit\"at des Saarlandes, D-66123 Saarbr\"ucken, Germany}
\address{Quantum Systems Unit, Okinawa Institute of Science and Technology Graduate University, Okinawa, 904-0495, Japan}
\author{Steve Campbell}
\address{Centre for Theoretical Atomic, Molecular, and Optical Physics, School of Mathematics and Physics, Queen's University Belfast, BT7 1NN, United Kingdom}
\author{Thomas Busch}
\address{Quantum Systems Unit, Okinawa Institute of Science and Technology Graduate University, Okinawa, 904-0495, Japan}
\author{Mauro Paternostro}
\address{Centre for Theoretical Atomic, Molecular, and Optical Physics, School of Mathematics and Physics, Queen's University Belfast, BT7 1NN, United Kingdom}

\begin{abstract}
We apply the framework of non-equilibrium quantum thermodynamics to the physics of quenched small-sized bosonic quantum gases in a one-dimensional harmonic trap.   
We show that dynamical orthogonality can occur in these few-body systems with strong interactions after a quench and we find its occurrence analytically for an infinitely repulsive pair of atoms. We further show this phenomena is related to the fundamental excitations that dictate the dynamics from the spectral function. We establish a clear qualitative link between the amount of (irreversible) work performed on the system and the establishment of entanglement. We extend our analysis to multipartite systems by examining the case of three trapped atoms. We show the initial (pre-quench) interactions play a vital role in determining the dynamical features, while the qualitative features of the two particle case appear to remain valid. Finally, we propose the use of the atomic density profile as a readily accessible indicator of the non-equilibrium properties of the systems in question.
\end{abstract}
\maketitle

\fancyhf{}
\rhead{\thepage}

\section{Introduction}
\label{intro}

Quantum gases offer a valuable platform for the study of quantum phenomena in interacting many-body systems. The availability of high-quality and reliable experimental control techniques and the low-level of external influences makes such systems excellent candidates for the simulation of quantum processes~\cite{Georgescu} and the exploration of the interplay between these and quantum critical behaviours.

Recent experimental progress has shown the possibility to observe non-equilibrium physics~\cite{Trotzky11}, which is quickly leading to the establishment of an experimental ultracold-atom framework for the exploration of complex phenomena, such as many-body localisation~\cite{Schreiber15}. 
In particular, this framework can be used to test the recently developed ideas for the finite-time thermodynamics of closed quantum systems~\cite{campisi}, which include tools for the quantification of thermodynamically relevant quantities, such as work and entropy, after a finite-time, non-equilibrium quantum quench. To date, this powerful formalism of non-equilibrium quantum thermodynamics has found only limited experimental validation and has been mostly applied to nuclear magnetic resonance settings~\cite{batalhao}. However, notwithstanding the exquisite control available over such systems, they are hard to scale and offer only few possibilities for the inclusion of many-body effects. On the contrary, ultracold atomic gases offer solutions for both of these issues and we will therefore in the following study the connections between the phenomenology of non-equilibrium quantum gases and finite-time thermodynamics.

For this, we consider small-size gases of interacting bosonic atoms that are perturbed out of their equilibrium configuration. In particular, we focus on the ground state of interacting, harmonically trapped bosonic atoms in one dimension, and subject them to a sudden quench of the Hamiltonian parameters. Such a perturbation, which has been recently explored to characterise the occurrence of Anderson's orthogonality catastrophe~\cite{Goold11, sindonaPRL,Campbell:14}, embodies the paradigm of a non-equilibrium process and has been shown to capture perfectly the complexity arising from quantum many-body effects in quantum spin systems~\cite{Dorner12}. In Ref.~\cite{sindonaNJP}, the case of a fermionic system was addressed and developed, the study of the statistics of work in bosonic Josephson junctions was presented in Ref.~\cite{lena}, and the study of quenched attractive cold gases has recently been examined in Ref.~\cite{piroli:16}. Furthermore, Ref.~\cite{adolfo} proposed realizing many-particle thermal machines using harmonically trapped bosons where a quantum advantage can be  achieved.

Here we explore the non-equilibrium properties of these trapped ultracold atoms. First, we find the analytical expressions for relevant thermodynamical quantities, such as the Loschmidt Echo (LE) and the average work, for a single atom subjected to a quench in the trapping frequency and elucidate the relation between the entropy production and the tendency toward dynamical orthogonality, although we find that the evolved state is never completely orthogonal to the initial one for finite quenches in the trapping frequency. Enlarging the system, we study the role particle interactions play. For the infinitely repulsive Tonks-Girardeu molecule we find analytical expressions for the same figures of merit as in the single atom case. While the qualitative behaviour is consistent with the single atom case, we now see due to the strong interactions, finite sized quenches lead to dynamical orthogonality. For finite interactions, we find numerically that the average in time of the entanglement and the (irreversible) average work are linked. This is a crucial result in this paper, as we find that even in the smallest interacting system, there exists a link between the establishment of correlations (entanglement) and the work done. We relate these features to the spectral function, which gives the fundamental excitations that dictate the dynamics. We complete our work studying the smallest possible mixture of ultracold bosons with both inter- and intra-species interactions: two atoms of type X and one atom of type Y and employ the same figures of merit as in the previous cases. We show a relationship between the  LE and the density profile, which is an experimentally measurable quantity. Furthermore this gives information about the correlations established among the atoms and indicates a trend for larger gases subjected to a quench, a situation of great theoretical interest and experimental relevance~\cite{Trotzky11}. Finally, we remark here that, using symmetry arguments, we show that the mixture of three atoms behaves identically to that of the three indistinguishable-boson system if both the inter- and intra-species constants are equal. 

Our presentation is organised as follows. In Section \ref{quantities} we provide a brief introduction to non-equilibrium quantum thermodynamics, focussing in particular on the consequences arising from a sudden quench.  This formal framework is then applied in Sections \ref{one}-\ref{three} to  the single atom, trapped molecule and three atom mixture when subjected to the quench of their Hamiltonian parameters. Finally, Section \ref{conclusion} draws our conclusions, while a set of technical considerations and details are presented in the Appendices.

\section{Non-equilibrium thermodynamics of quantum quenches}
\label{quantities}

In the following we will first briefly summarise the key notions of finite-time thermodynamics in closed quantum systems.
For this we consider a system whose Hamiltonian, ${\cal H}$, depends on an externally controlled, time-dependent {\it work parameter} $\lambda_t$. The system is assumed to be in contact with a bath at inverse temperature $\beta$ for a time long enough to have reached equilibrium. At $t=0$, the system is detached from the reservoir and its energy is changed by modifying the value of the work parameter from $\lambda_0$ to $\lambda_\tau$. The evolution is accounted for by the unitary propagator $U_\tau$. As the system is detached from the surrounding world, such a change of energy can only be interpreted as work done on/by the system, which can be characterised by introducing the work probability distribution~\cite{campisi}
\begin{equation}
P(W)=\sum_{n,M} p(n,M) \delta\left[W-(E'_M-E_n)\right].
\label{eq:qworkdist}
\end{equation}
Here $E_n$ ($E'_M$) is the $n^{\rm th}$ ($M^{\rm th}$) eigenvalue of the associated eigenstate $\ket{n}$ ($\ket{M}$), of the initial (final) Hamiltonian. Moreover, $p(n,M)=\mathrm{Tr}[\ket{M}\bra{M} U_\tau \ket{n}\bra{n}\rho_s\ket{n}\bra{n} U^\dagger_\tau]$ is the joint probability of finding the system in $\ket{n}$ at time $t=0$ and in state $\ket{M}$ at time $\tau$, after the evolution by the time-propagator $U_\tau$. Obviously, such a joint probability can be decomposed as $p(n,M)=p^0_n\;  p^\tau_{M \vert n}$, where $p_n^0$ is the probability that the system is found in state $\ket{n}$ at time $t=0$ and $p^\tau_{M\vert n}$ is the conditional probability to find the system in $\ket{M}$ at time $\tau$ if it was initially in $\ket n$. Therefore, $P(W)$ contains information about the statistics of the initial state and the fluctuations arising from quantum dynamics and measurement statistics. The characteristic function of the work probability distribution of $P(W)$ is defined as~\cite{campisi}
\begin{equation}
\chi(u,\tau)=\int\!{dW}e^{iuW}P(W)={\rm Tr}[{{U}^\dag_\tau e^{iu{\cal H}(\tau)} U_\tau e^{-iu{\cal H}(0)}\rho^0_{\rm eq}}],
\label{characteristicfunction}
\end{equation}
with $\rho^0_{\rm eq}$ being the initial equilibrium state of the system and ${\cal H}(\tau)$ the Hamiltonian of the system when the work parameter takes the value $\lambda_\tau$. 

For a quasistatic process, the change in free energy $\Delta F$ of the system is equal to the average work done on/by it. The former can be written as $\Delta F=\Delta E-\Delta S/\beta$ with $\Delta E$ being the change in energy and $\Delta S$ the corresponding entropy variation. On the other hand, if the process is fast (i.e. not quasistatic), then the relation $\langle W\rangle\ge\Delta F$ holds, accounting for the fact that part of the work performed on/by the system is {\it dissipated} due to the abrupt nature of the transformation. By introducing the standard definition of non-equilibrium entropy production 
\begin{equation}
\langle\Sigma\rangle=\Delta S-\beta \langle Q\rangle,
\end{equation}
where $\langle Q\rangle$ is the average heat exchanged with the environment, we find for a closed, unitarily evolving system 
\begin{equation}
\label{sigma}
\langle\Sigma\rangle=\beta(\langle W\rangle-\Delta F).
\end{equation}
This allows to quantify the irreversible nature of a given process in terms of the discrepancy between $\Delta F$ and $\langle W\rangle$. The definition of $\langle\Sigma\rangle$ allows for the consideration of the so-called irreversible work
\begin{equation}
\label{irrwork}
\big< W_{\rm irr} \big>=\big< W \big> -  \Delta F,
\end{equation}
which will be extensively used in this work. A general approach to irreversible entropy in open quantum systems (including non-equilibrium ones) can be found in Ref.~\cite{Deffner}, while a different quantifier which is based on the use of adiabatic transformations (rather than the implicit isothermal ones considered here) has been proposed in Ref.~\cite{Plastina}.

A very useful lower bound to the non-equilibrium entropy production, $\langle\Sigma\rangle$, can be based on the unitarily invariant Bures angle (see Ref.~\cite{deffnerPRE}). For arbitrary density matrices $\rho_{1,2}$, the Bures angle is defined as $\mathcal{B}=\arccos\left( \sqrt{F(\rho_1,\rho_2)} \right)$ with $F(\rho_1,\rho_2)$  the fidelity between the two states. Using this we find
\begin{equation}
\label{bound}
\langle \Sigma\rangle\ge\big< \Sigma \big>_B = \frac{8}{\pi^2} \mathcal{B}^2_{\rm eq}(\tau),
\end{equation}
where $\mathcal{B}_{\rm eq}(\tau)=\arccos\left( \sqrt{F(\rho^\tau_{\rm eq},\rho_\tau)} \right)$ is the angle between the non-equilibrium state $\rho_\tau$ of a closed quantum system, and its equilibrium version $\rho^\tau_{\rm eq}$. Eq.~(\ref{bound}) defines a thermodynamic distance that is valid arbitrarily far from equilibrium, and can thus be used to characterise the departure from equilibrium following an arbitrary driving process.

Since we are interested in examining the dynamics of a cold atomic system after a sudden quench, we will make use of the fact that we start from the ground state of a given system and that its state remains pure throughout the whole dynamics. We will use the spectral decomposition of the initial and final Hamiltonians of the system $\mathcal{H}_\alpha= \sum_j E_j^\alpha \ket{\psi_j^\alpha}\bra{\psi_j^\alpha}$ with $\alpha=I$ ($\alpha=F$) denoting the initial (final) Hamiltonian operator. Here $E^\alpha_j$ is the $j^{\rm th}$ eigenvalue of ${\cal H}_\alpha$ with associated eigenstate $\ket{\psi^\alpha_j}$. 

A key figure of merit for our system is the Loschmidt echo (LE), which is defined as 
\begin{eqnarray}
\label{EchoEq}
\mathcal{L}(t)&=\vert \bra{\Psi_0} e^{i\mathcal{H}_F t}e^{-i\mathcal{H}_I t}\ket{\Psi_0}\vert^2=\Big\vert  \sum_n e^{i(E_n^F-E_0^I)t} \big<\psi_0^I \vert\psi_n^F\big>^2  \Big\vert^2,
\end{eqnarray}
where we have assumed that the initial state of the system $\ket{\Psi_0}$ coincides with the ground state $\ket{\psi^I_0}$ of ${\cal H}_I$. The LE is closely related to the characteristic function of the probability distribution of the work done on/by the system upon subjecting it to the quench considered here. In fact, for a sudden quench we have that $U_{\tau}=1\!\!1$, with $1\!\!1$ the identity operator, and thus $\chi(u,\tau)\equiv\chi(\tau)={\rm Tr}[e^{iu{\cal H}(\tau)}e^{-iu{\cal H}(0)}\rho^0_{\rm eq}]$. Here, we are taking $\rho^0_{\rm eq}=\ket{\psi^I_0}\bra{\psi^I_0}$  and, by using the identifications ${\cal H}(0)={\cal H}_I$ and ${\cal H}(\tau)={\cal H}_F$, we find ${\cal L}(t)=|\chi(t)|^2$  with
\begin{equation}
\chi(t) =\bra{\psi^I_0} e^{i\mathcal{H}_F t}e^{-i\mathcal{H}_I t}\ket{\psi^I_0} = \sum_j e^{i(E_j^F -E_0^I)t} \big\vert \big<\psi_0^I \vert\psi_j^F\big> \big\vert^2.
\end{equation}
From this the average work is given by 
\begin{eqnarray}
\label{Wav}
\langle W \rangle = -i \partial_t \chi(t)  {\big|}_{t=0} &=
\sum_j (E_j^F -E_0^I) \big\vert \big<\psi_0^I \vert\psi_j^F\big> \big\vert^2,
\end{eqnarray}
while from the definition of irreversible entropy production Eq.~(\ref{sigma}), we can introduce a quantifier of the dissipated work due to the non-quasistatic nature of the quench. 

\section{Single trapped atom}
\label{one}
Let us start by considering the simplest possible scenario of a harmonically trapped single  atom in one dimension. The Hamiltonian of the system reads
\begin{equation}
{\cal H}=-\frac{\hbar^2}{2m}\partial^2_x+\frac12 m\omega^2_1x^2
\end{equation}
with $m$ the mass of the atom and $\omega_1$ the frequency of the trapping potential. In the following we will consider a quench in the trapping potential frequency  $\omega_1\to\omega_2$ and, in order to simplify our notation, we rescale the position of the atom with respect to the ground state length $a_{{\rm ho}}=\sqrt{\hbar/m\omega_2}$, and its energy with respect to $\hbar\omega_2$. The dimensionless initial Hamiltonian $\tilde{\cal H}={\cal H}/(\hbar\omega_2)$ of the system then  reads
\begin{equation}
\tilde{\cal H}^I=-\frac12\partial^2_{\tilde x}+\frac{{\tilde x}^2}{2\epsilon^2}
\label{eq:InitialHamiltonian}
\end{equation}
where $\tilde{x}=x/a_{\rm{ho}}$ and $\epsilon=\omega_2/\omega_1$. 

\begin{figure}[t!]
\begin{center}{\bf (a)}\hskip0.3\columnwidth {\bf (b)}\hskip0.3\columnwidth {\bf (c)}\end{center}
\includegraphics[width=0.33\columnwidth]{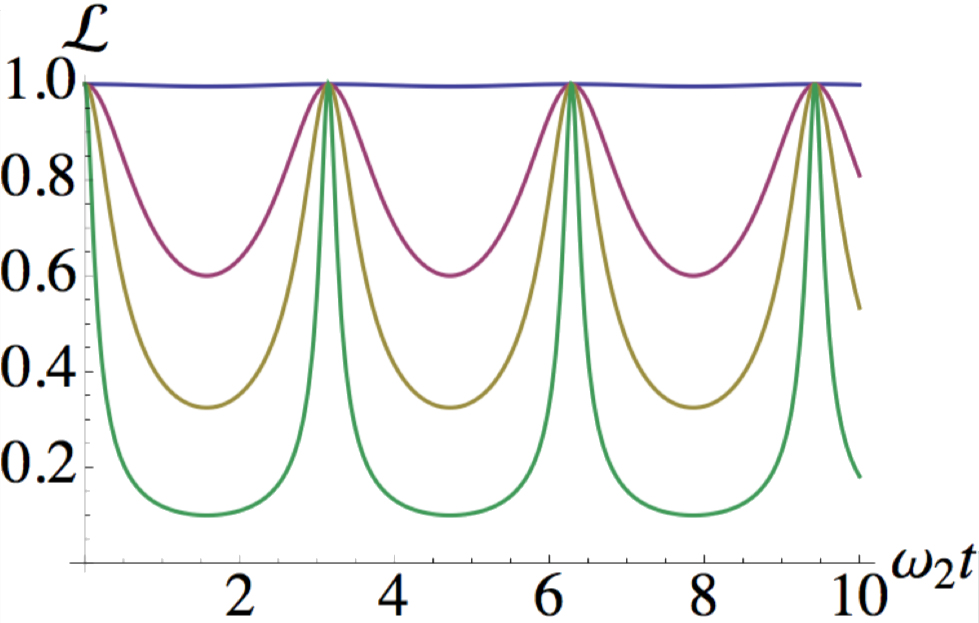}~~\includegraphics[width=0.33\columnwidth]{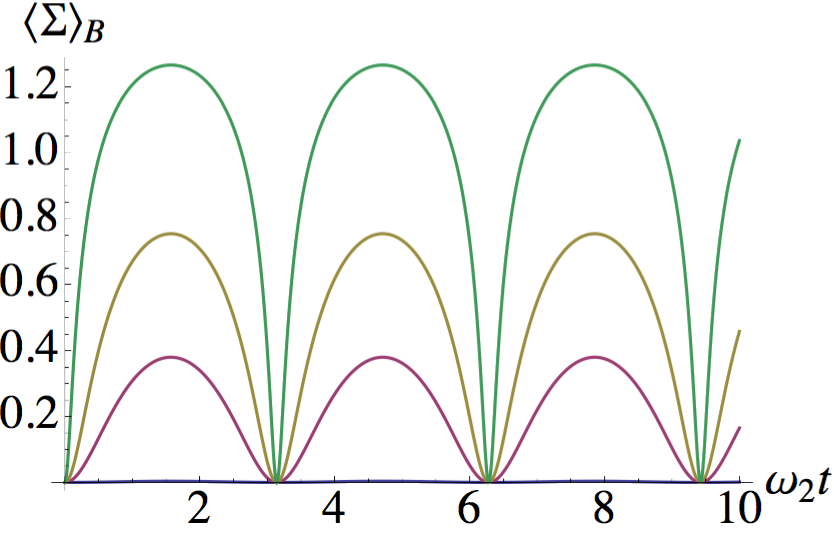}~~\includegraphics[width=0.33\columnwidth]{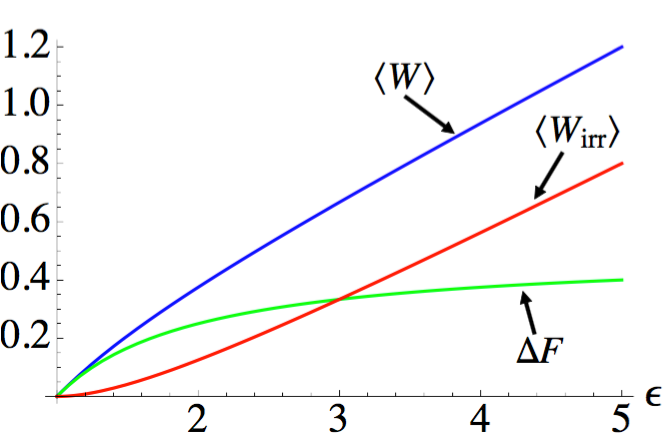}\\
\caption{{\bf (a)} The LE for a single trapped atom for increasingly large quenches $\epsilon=1.1$, 3, 6, 20 (top to bottom). {\bf (b)} Lower bound on the irreversible entropy produced by a sudden quench of the trap frequency with $\epsilon=1.1$, 3, 6, 20 (bottom to top). {\bf (c)} Average work (blue), irreversible work (red) and free energy (green) for a single trapped atom.}
\label{fig1n}
\end{figure}
The results presented in this section are closely related to those presented in Refs.~\cite{sindonaNJP,deffnerPRE}. However an explicit re-examination of these calculations will be useful for understanding the upcoming sections. To calculate any of the above quantities requires determining the overlap between the initial (ground) state and the eigenstates of $\tilde{\cal H}^F=-\frac12\partial^2_{\tilde x}+\frac{{\tilde x}^2}{2}$ [cf. Eq.~(\ref{Wav})] which in this case is done using the known wavefunctions
\begin{equation}
\psi_0^I(\tilde x) =\sqrt[4]{\frac{1}{\pi\epsilon}} e^{-\frac{1}{2\epsilon}{\tilde x}^2}\quad\textrm{and}\quad 
\psi_n^F(\tilde x) =\sqrt[4]{\frac{1}{\pi}} \frac{e^{-\frac{1}{2}{\tilde x}^2}}{\sqrt{2^n n!}} H_n\left( \tilde x \right),
\end{equation}
with the associated energies $E_0^I=1/(2\epsilon)$ and $E_n^F=(n+{1}/{2})$, and where $H_n(y)$ is the Hermite polynomial of order $n$ and argument $y$.
Exploiting the fact that $\langle{\psi^I_0}\ket{\psi^F_{2k+1}}=0~(k\in{\mathbb N}^{0})$, we find 
\begin{equation}
\label{singleoverlap}
\big<\psi_0^I \vert\psi_n^F\big> =
 \left(\frac{2\sqrt{\epsilon}}{n! (\epsilon+1)}\right)^\frac{1}{2} \left(\frac{\epsilon-1}{\epsilon+1}\right)^{\frac{n}{2}} (n-1)!!,
\end{equation}                
which is valid only for even values of $n$ and directly leads to
\begin{eqnarray}
\mathcal{L}(t)&=\frac{2\epsilon}{\sqrt{\left[ 2\epsilon\cos\left( \omega_2 t \right) \right]^2+\left[ \left( 1+\epsilon^2 \right)\sin\left(\omega_2 t  \right) \right]^2}},\label{eq:LE}\\
\langle W \rangle &=\frac{\epsilon^2-1}{4\epsilon}.\label{cooleq}
\end{eqnarray}
Note that the average work  is dimensionless in our chosen units. As our system is pure, the free energy difference is simply the difference between the initial and final ground state energies, $\Delta F=\frac{1}{2\epsilon}(\epsilon-1)$, and thus
\begin{equation}
\langle W_{\rm irr} \rangle = \frac{(\epsilon-1)^2}{4\epsilon}.
\end{equation}
The lower bound of the entropy produced dynamically is given by
\begin{equation}
\big< \Sigma \big>_B = \frac{8}{\pi^2} \left[ \arccos\frac{2\sqrt{\epsilon}}{\sqrt[4]{ (1-\epsilon)^4+(1+\epsilon)^4-2(1-\epsilon^2)^2 \cos(2\omega_2 t)}} \right]^2,
\end{equation}
and in Fig.~\ref{fig1n} we show the behaviour of these quantities for different representative values of the quench. Examining panels {\bf (a)} and {\bf (b)} we see an oscillating pattern stemming from the harmonic oscillator dynamics and find that the behaviour of the lower bound on the irreversible entropy is strongly correlated with the behaviour of the LE. Its value at a given time grows with the strength of the quench as a consequence of the fact that, as $\epsilon$ grows, the ground state of $\tilde{\cal H}_F$ becomes increasingly different from $\ket{\psi^I_0}$. When examined against time, we find that the maximum entropy production is achieved in correspondence with the minimum value of ${\cal L}(t)$. At this time, the state of the system is as different as possible from the the initial one, which coincides with the maximum irreversible entropy generated. By inspection of Eq.~(\ref{eq:LE}), we note that full dynamical orthogonality never occurs for finite quenches of the trapping frequency. In Fig.~\ref{fig1n} {\bf (c)} we show the behavior of the average work, irreversible work and the free energy against the strength of the quench. While, naturally, all quantities grow with increasingly large quenching strengths, $\big<W_{\rm irr}\big>$ grows much more significantly than $\Delta F$, precisely inline with the increasingly large maxima attained in the entropy produced for larger quenches.

\section{Trapped molecule}
\label{molecule}
We now move to a more complex multi-particle system and examine the effect of atomic interactions on the quantities discussed above. For this we consider two atoms of equal mass $m$, jointly trapped in a harmonic potential of frequency $\omega_1$ and then quench the trap frequency to  $\omega_2$, as before. The Hamiltonian model of the system scaled in the same way as in Eq.~(\ref{eq:InitialHamiltonian}) reads
\begin{equation}
\label{ham}
{\cal H}^{I,F}=\sum_{j=1,2}\left[-\frac12\partial^2_{\tilde x_j}+\frac{\tilde x_j^2}{2}\frac{\omega_k^2}{\omega_2^2}\right]+ g\delta(\tilde x_1-\tilde x_2),
\end{equation}
and corresponds to the initial Hamiltonian for $k=1$ the the final one for $k=2$. The parameter $g$ is the coupling constant, which characterises the boson-boson interaction in the limit of low temperatures. We will assume the interactions to be repulsive throughout this work and therefore $g$ to be positive.  
Note that the rescaling leads to a coupling constant in units of  $a_{\mathrm {ho}}\hbar\omega_2$ and this again allows to define the parameter $\epsilon=\omega_2/\omega_1$. 

It is important to note that while in principle we have the freedom to quench the interaction strength or the trapping frequency, quenching $\omega$ will implicitly lead to a quench in the interaction strength, if it was initially finite. We will show below that this can be dealt with cleanly for two-particle systems, however for larger system sizes more care must be taken (see Sec.~\ref{three}).

\subsection{Tonks-Girardeau pair of atoms}
\begin{figure}[t!]
\begin{center}
{\bf (a)}\hskip0.5\columnwidth{\bf (b)}\\
\includegraphics[width=0.45\columnwidth]{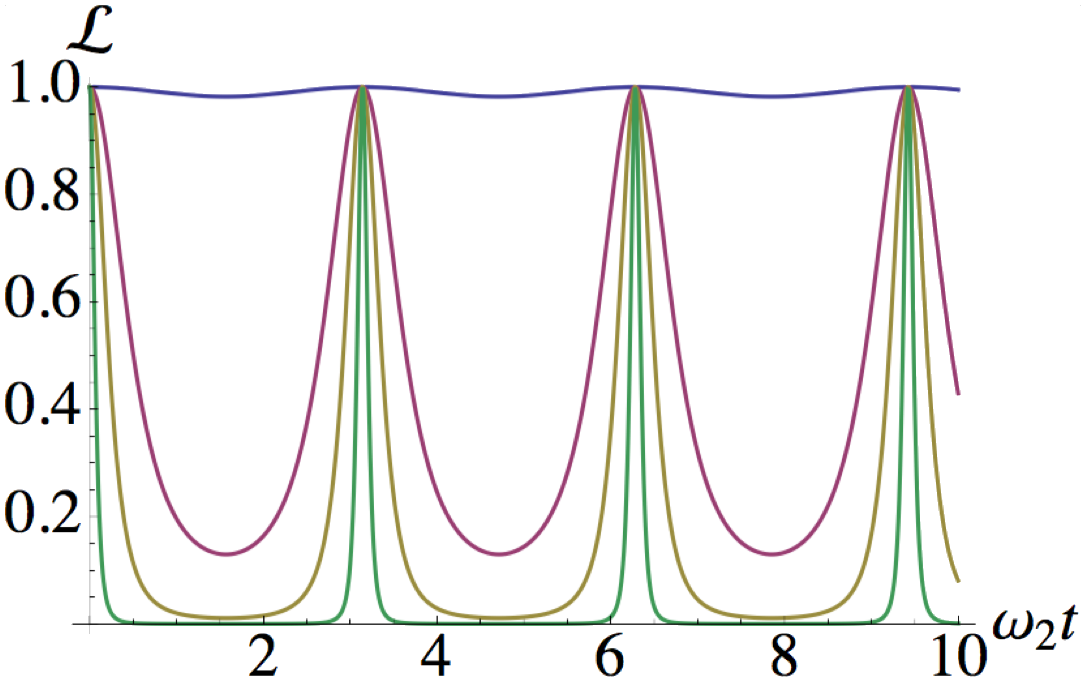}~~~\includegraphics[width=0.45\columnwidth]{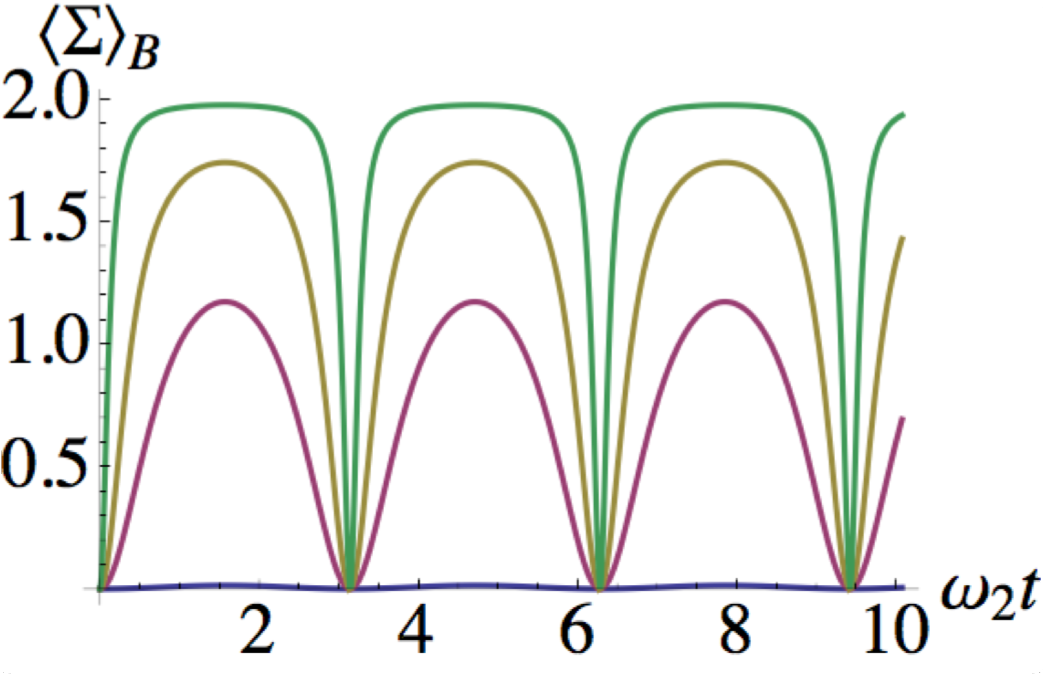}
\caption{{\bf (a)} Temporal behaviour of the LE of the TG molecule for increasingly large quenches $\epsilon=1.1, 3, 6, 20$ (from top to bottom curve, respectively). {\bf (b)} Lower bound on the irreversible entropy produced through the quench. We have used the same quenching amplitudes and colour code as in panel {\bf (a)}.}
\label{two}
\end{center}
\end{figure}

The coupling constant can range from zero to infinity, which is the so-called Tonks-Girardeau (TG) limit. Since in the non-interacting case all the results of the previous section still apply, we will in the following start by carefully studying the TG limit, where the atoms behave as hardcore bosons and are readily amenable to analytic treatment. The wavefunction of the system can be split into its centre-of-mass (COM) and relative (REL) coordinates, $\ket{\psi^\alpha_n}=\ket{\eta^\alpha_n}\ket{\varphi^\alpha_n}$, where $\ket{\eta^\alpha_n}$ ($\ket{\varphi^\alpha_n}$) refers to the COM (REL) degree of freedom. The LE and characteristic function depend on the overlap between the initial and final wavefunctions, as before. In fact, the wavefunctions of the COM terms are precisely the same as in the single atom problem and therefore the overlap is given by Eq.~(\ref{singleoverlap}). However, due to the infinite interaction each even REL state becomes degenerate with  the next higher lying odd state, such that it is sufficient to  work only with the odd states.  The required initial and final eigenstates are 
\begin{eqnarray}
&\varphi_1^I(\tilde x) = \left(\frac{4}{\pi}\right)^{\frac{1}{4}}\frac{1}{\epsilon}  \tilde x e^{-\frac{1}{2\epsilon}\tilde x^2},\,\nonumber\\
&\varphi_{2n+1}^F(\tilde x) = \sqrt{\frac{1}{2^{2n+1}(2n+1)!}}
\left(\frac{1}{\pi}\right)^{\frac{1}{4}} e^{-\frac{\tilde x^2}{2}} H_{2n+1}\left( \left\vert \tilde x \right\vert \right),
\end{eqnarray}
where $\tilde x = \tilde x_1 - \tilde x_2$, with the associated energies $A_1^I={3}/2\epsilon$ and $A_{2n+1}^F=\left(2n+{3}/{2}\right)$. We can then express the average work and the LE in terms of these functions as
\begin{equation}
\label{tonksLE}
\mathcal{L}(t)= \left\vert \sum_{n,p} e^{i( E_{2n}^F+A_{2p+1}^F-E_0^I-A_1^F )t} \vert \big< \eta_0^I | \eta_{2n}^F \big>\big< \varphi_1^I | \varphi_{2p+1}^F \big>  \vert^2 \right\vert^2,
\end{equation}
\begin{equation}
\big<W\big>=\sum_{n,p} ( E_{2n}^F+A_{2p+1}^F-E_0^I-A_1^F ) \vert \big< \eta_0^I | \eta_{2n}^F \big>\big< \varphi_1^I | \varphi_{2p+1}^F \big>  \vert^2,
\end{equation}
with the overlap between the REL wavefunctions given by 
\begin{equation}
\big< \varphi_1^I | \varphi_p^F \big> = \frac{ 2^{\frac{p+3}{2}}\sqrt[4]{(1/\epsilon)^3} p!!}{\sqrt{2^{p} p! (1/\epsilon+1)^{p+2}}} \left(1-\frac{1}{\epsilon}\right)^{\frac{p-1}{2}}.
\end{equation}
Using this expression we can calculate the LE using Eq.~(\ref{tonksLE}) and $\big<\Sigma\big>_B$ using Eq.~(\ref{bound}). The infinite sums can be evaluated explicitly, and the final expression for the average work turns out to be exactly four times the single-atom work
\begin{equation}
\big< W \big> = \frac{\epsilon^2-1}{\epsilon},
\end{equation}
showing that in this regime work is an extensive quantity as we have performed a quench of the trap frequency for two atoms [compare to Eq.~(\ref{cooleq})]\footnote{In the TG limit we find the average work scales as $N^2$ times the single atom average work.}. This can be further  understood from the fact that in the TG limit the bosons behave as non-interacting fermions~\cite{sindonaNJP}. In Fig.~\ref{two} we show the behaviour of $\mathcal{L}(t)$ and $\big<\Sigma\big>_B$. While the qualitative behaviour is consistent with the single atom case, we see that the effect of the interactions is to magnify these features as we now must account for many-body effects. In particular we also see the system periodically evolves into orthogonal states as the interacting two-body system can be moved further out of equilibrium, as evidenced by the vanishing values of LE which are already less than $10^{-2}$ for $\epsilon=6$.

\subsection{Finitely interacting pair of atoms}
When the frequency of the trap is quenched while the interaction strength is finite, i.e.~between the non-interacting and the TG limit, the resulting complexity requires we use a numerical approach to study the system, and we refer the reader to Refs.~\cite{Garcia-march:12,Garcia-march:13} for details of the recipes employed here. We remark that one could equally employ the known analytic solutions in Ref.~\cite{Busch:98}, however as the infinite sums and overlaps must still be performed numerically such an approach is equivalent to the one employed here. The dynamics well characterised by the von Neumann entropy (vNE) of the atomic state, which is given by
\begin{equation}
 S=-\mbox{Tr}\left[\rho\ln\rho\right],
\end{equation}
where $\rho$ is the reduced density matrix (RDM) for one of the atoms. Since we consider the system to be always in a pure state, the vNE is a good measure of the entanglement in the system.
\begin{figure}[h!]
\begin{center}
{\bf (a)}\hskip0.45\columnwidth{\bf (b)}
\includegraphics[width=0.45\columnwidth]{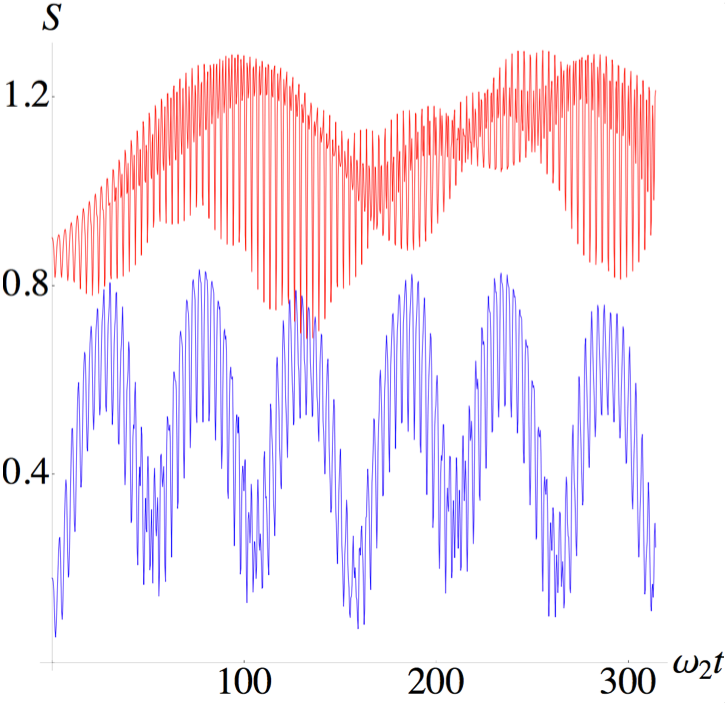}~\includegraphics[width=0.45\columnwidth]{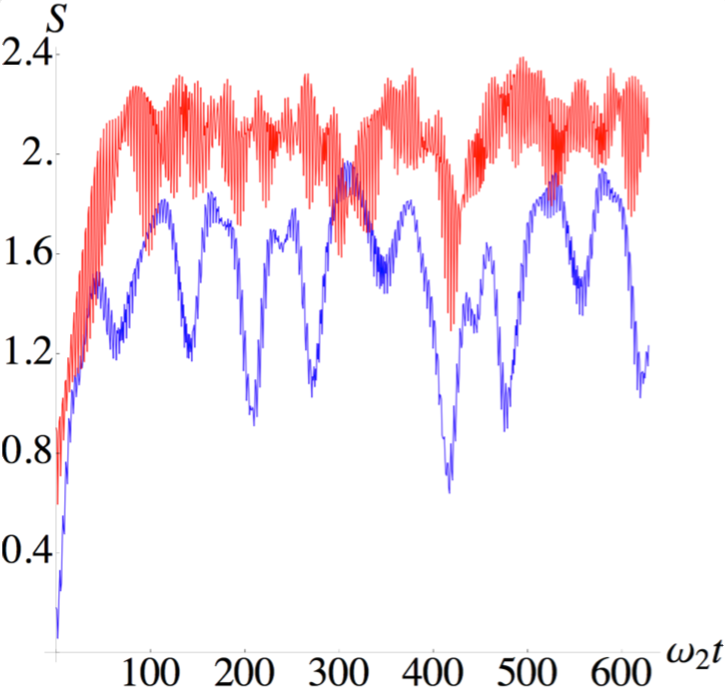}
\caption{ {\bf (a)} Evolution of the vNE following a quench of the trap frequency against the dimensionless time $\omega_2 t$ for $\epsilon=2$ and two values of the coupling strength. The red (blue) curve is for $g=20$ ($g=1$). {\bf (b)} Same as panel {\bf (a)} with $\epsilon=5$.}
\label{mossy1}
\end{center}
\end{figure}
\begin{figure}[h!]
\begin{center}
{\bf (a)}\hskip0.35\columnwidth{\bf (b)}
\includegraphics[width=0.35\columnwidth]{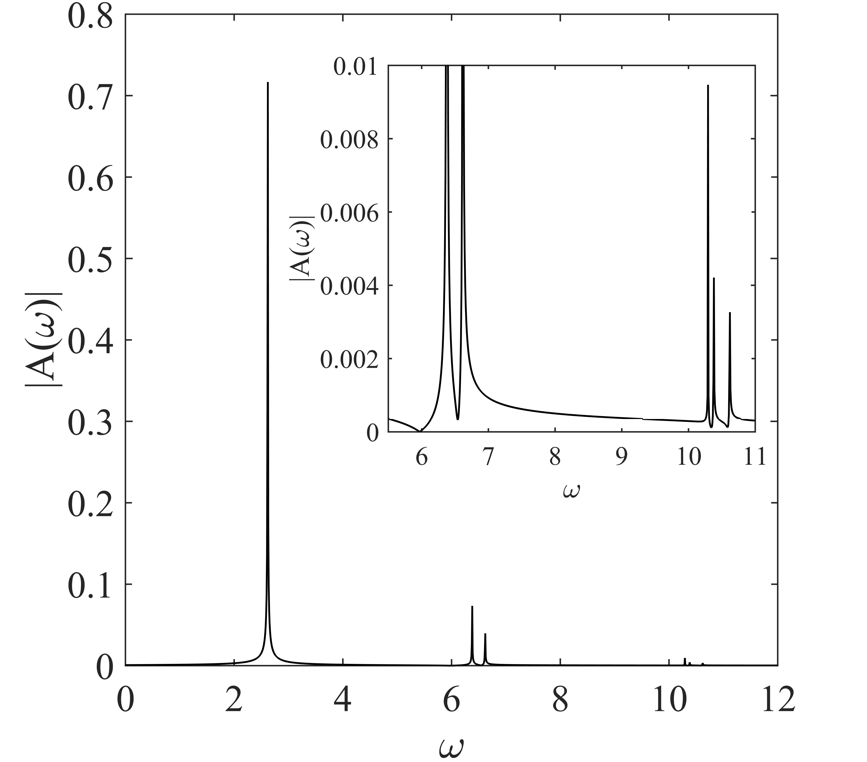}~~\includegraphics[width=0.32\columnwidth]{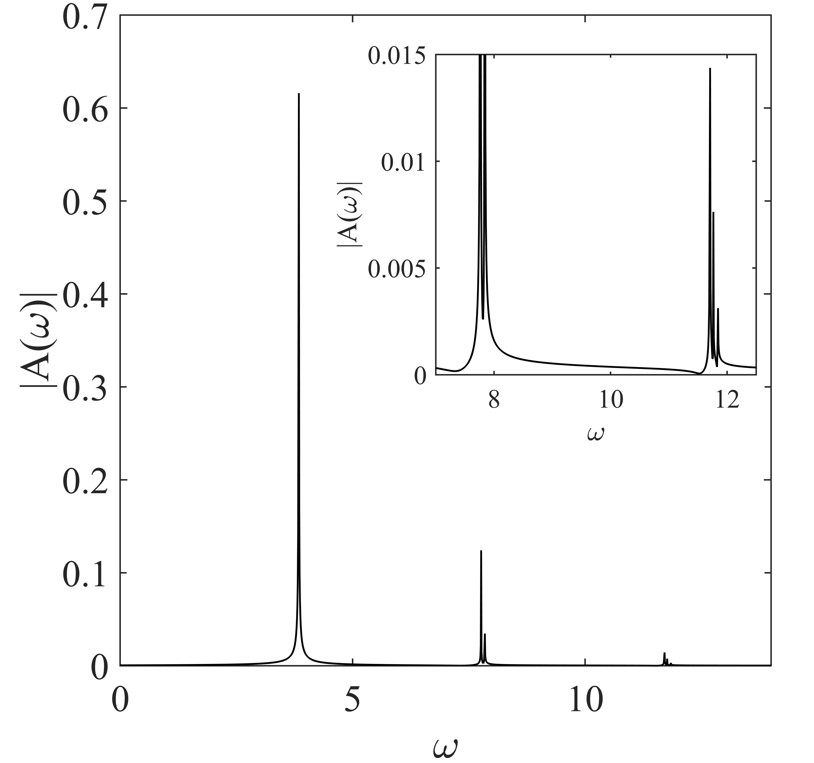}\\
{\bf (c)}\hskip0.35\columnwidth{\bf (d)}
\includegraphics[width=0.35\columnwidth]{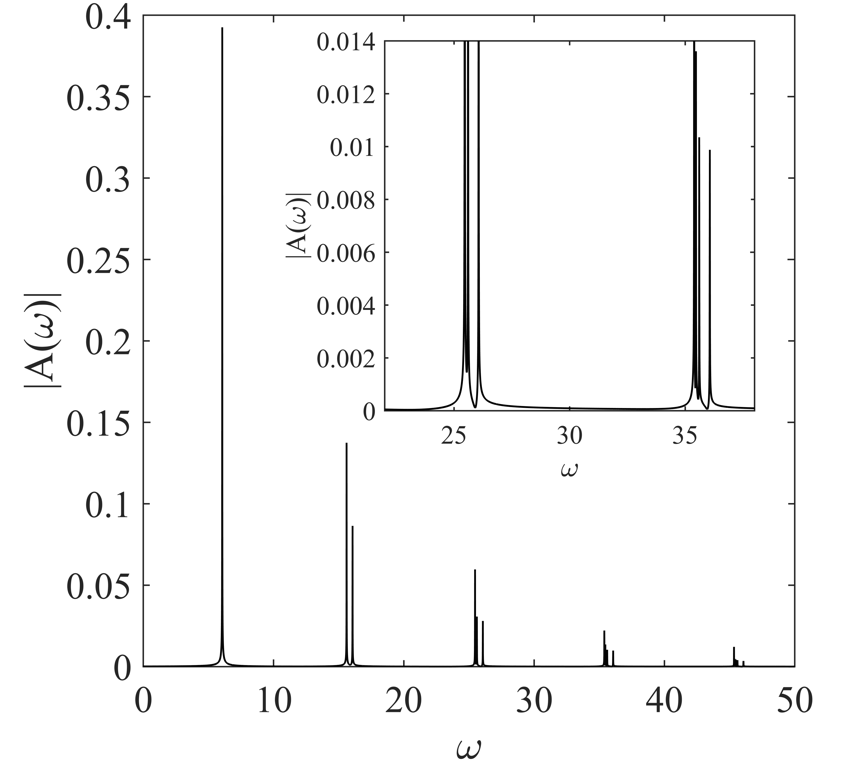}~~\includegraphics[width=0.35\columnwidth]{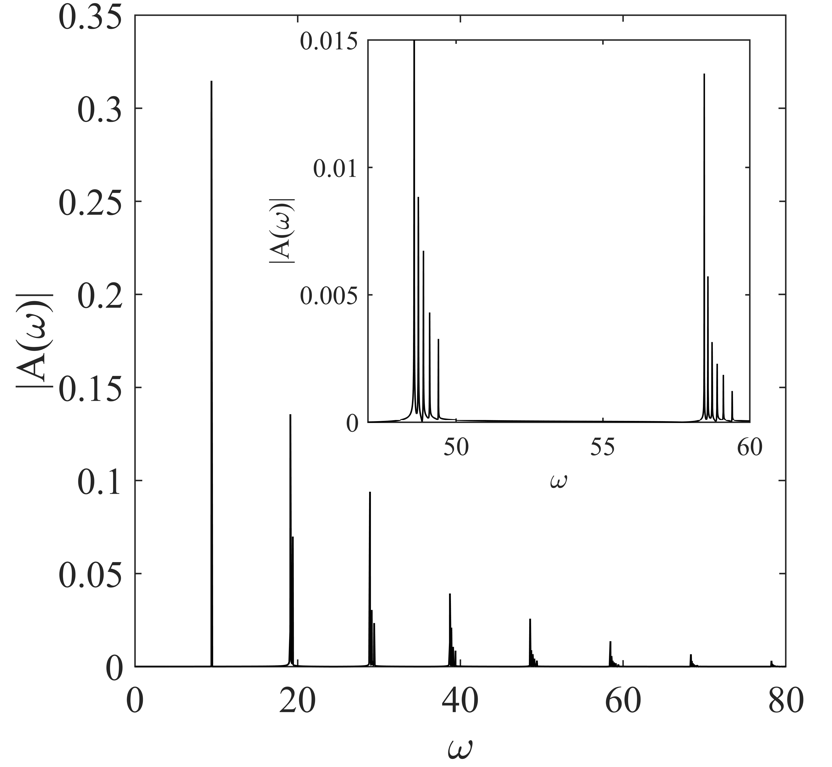}\\
\caption{Spectral function for $\epsilon=2$ with {\bf (a)} $g=1$ and {\bf (b)} $g=20$, and for the stronger quench of $\epsilon=5$ with {\bf (c)} $g=1$ and {\bf (d)} $g=20$. }
\label{spec}
\end{center}
\end{figure}
For finite values of $g$, quenching the trap frequency implies also a quench of the interaction strength between the atoms and the evolution of the vNE will therefore be determined by a competition between these two mechanisms.\footnote{In the TG limit the vNE is constant for all quenches, due to the infinite value of the interaction. We remark $S=0.68275$ (two particles) and $S=1.0574$ (three particles).} The resulting behaviour is shown in Fig.~\ref{mossy1} for various values of the quench amplitude and the coupling strength $g$. For small-amplitude quenches and weakly interacting atoms (lower blue curve in Fig.~\ref{mossy1}{\bf (a)}) the vNE oscillates, as expected for a quench in a harmonic oscillator, with an amplitude modulation due to the effective interaction quench. At larger strengths of the quench (lower blue curve of panel {\bf (b)})  this behaviour is strongly modified. The absolute values of the entanglement increase, as the system becomes more strongly correlated, but there is no longer evidence of regular oscillations as the spectrum has become anharmonic due to the interactions. Looking at the spectral function of the out-of-equilibrium state, $A(\omega)=2 \mbox{Re} \int \chi(t) e^{i\omega t} dt$~\cite{Mahan}, we can identify these different excitation frequencies inherent in the evolution, see Fig.~\ref{spec}. The majority of the motion is governed by the quasi particle peak at the ground state energy of the quenched state $E_0^F+A_0^F$, and smaller contributions stem from combinations of COM and REL even states at higher energies (there are no contributions from the odd states). At larger interaction strengths the high energy peaks in the spectral function approach each other, as the system becomes doubly degenerate in the TG limit, see Fig.~\ref{spec} {\bf (b)}. This causes larger interference effects that are apparent in the entropy evolution (c.f. the upper red curves in Fig.~\ref{mossy1} {\bf(a)} and {\bf(b)}). The periodic nature of the revivals is destroyed for the large quench of $\epsilon=5$ due to the broadening of the high energy peaks (see inset of Figs.~\ref{spec} {\bf(c)} and {\bf(d)}) which highlights the chaotic dynamics of the strongly quenched state. A clear signature of interference of the COM and REL states is manifested in the appearance of Fano-resonances in the spectral function, a feature which is typical in systems with two scattering amplitudes which overlap (see all insets in Fig.~\ref{spec})~\cite{Fano}.

\begin{figure}[t]
\begin{center}
{\bf (a)}\hskip0.35\columnwidth{\bf (b)}
\includegraphics[width=0.4\columnwidth]{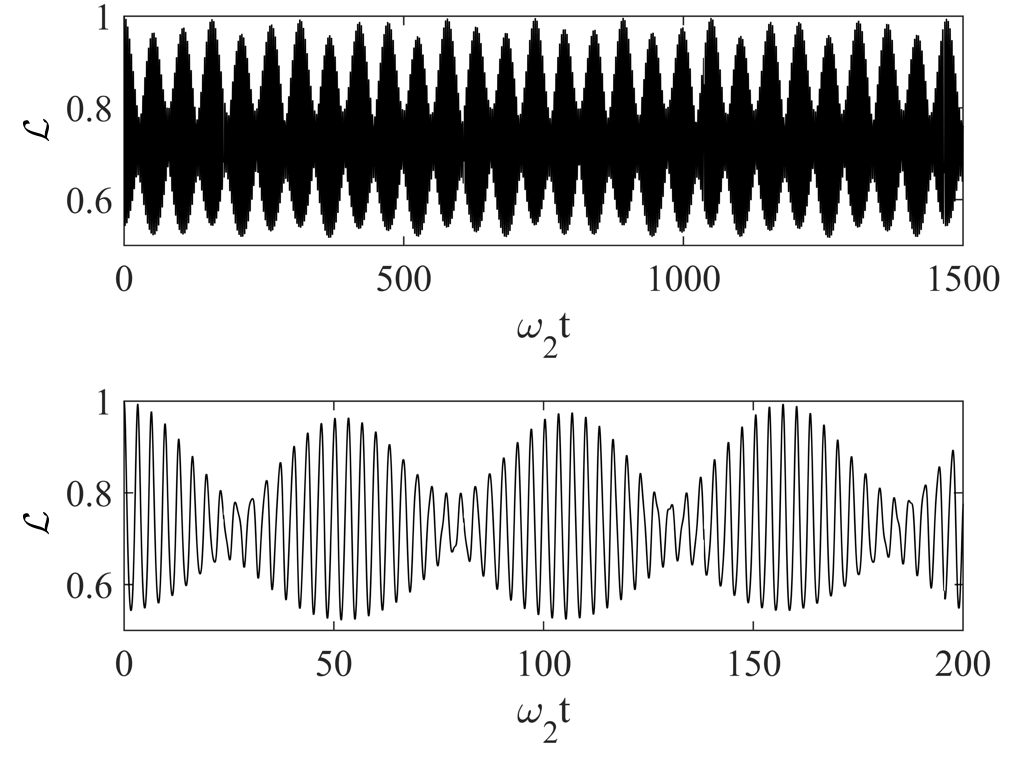}~~\includegraphics[width=0.4\columnwidth]{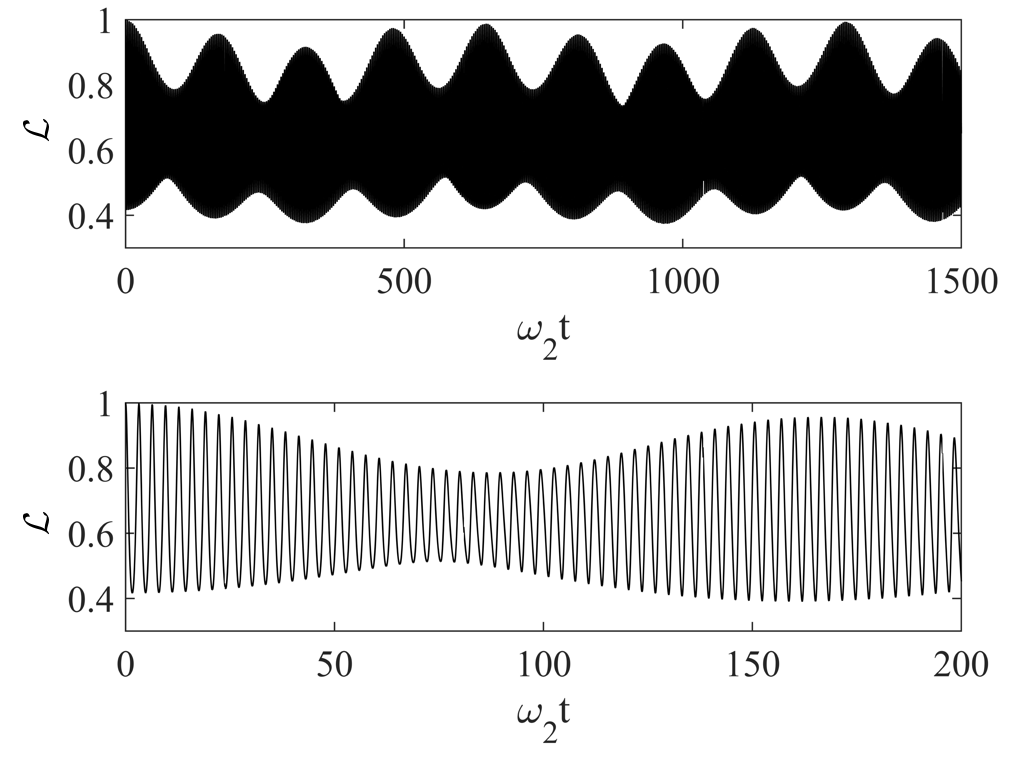}\\
{\bf (c)}\hskip0.35\columnwidth{\bf (d)}
\includegraphics[width=0.4\columnwidth]{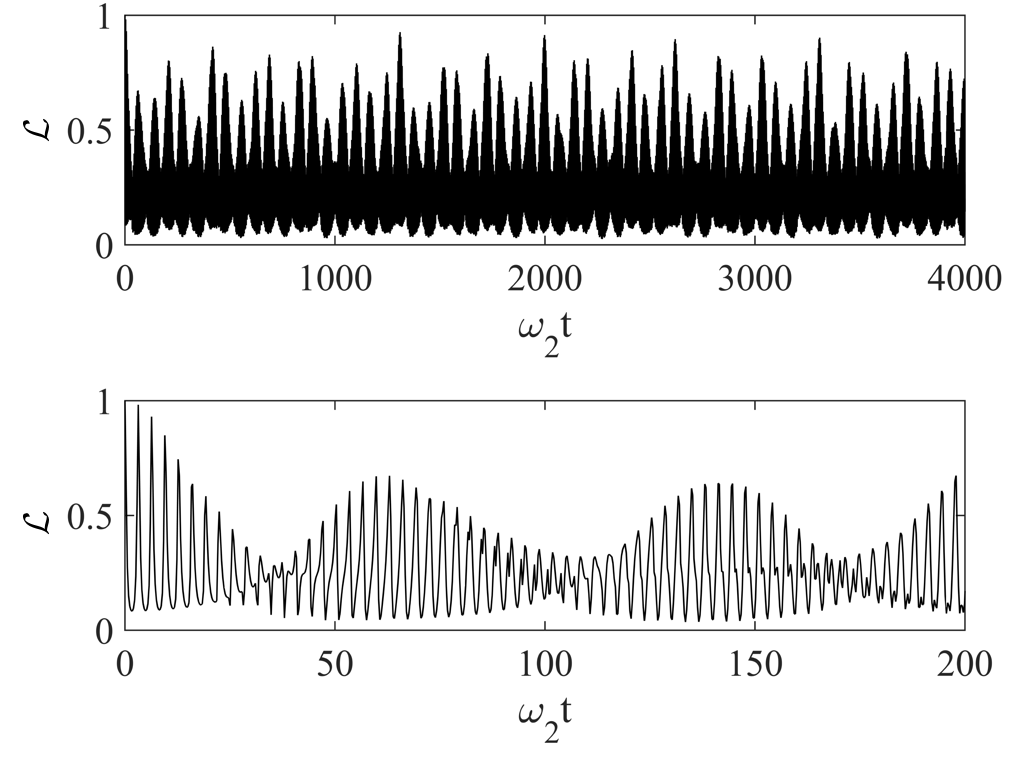}~~\includegraphics[width=0.4\columnwidth]{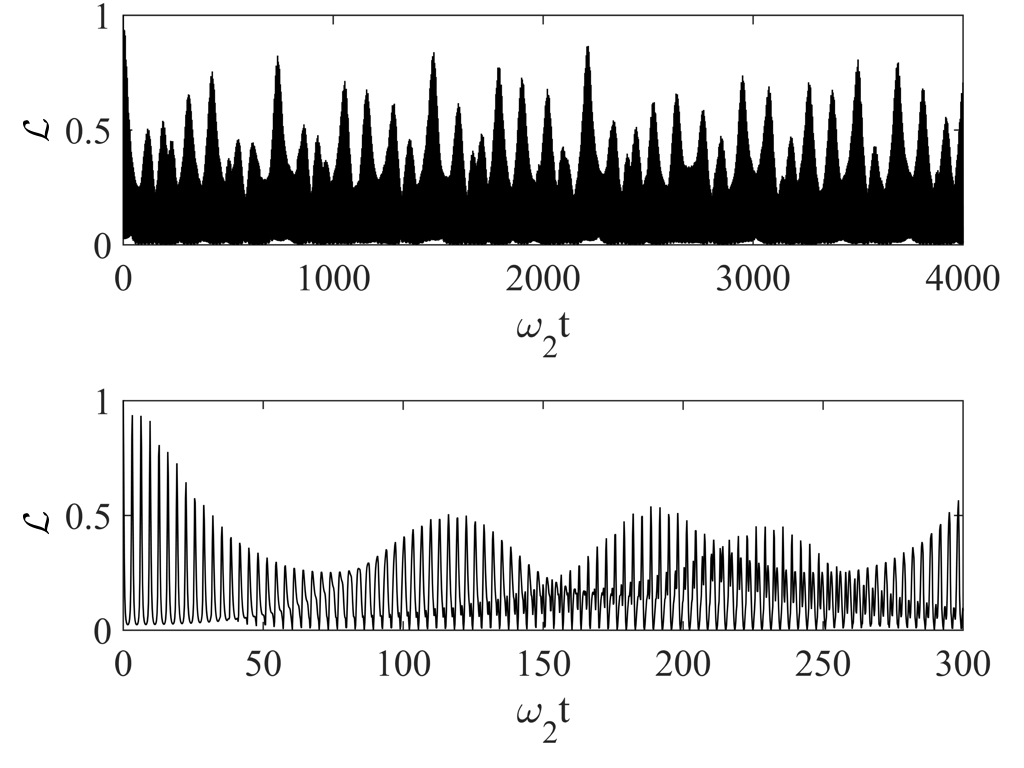}\\
\caption{LE for $\epsilon=2$ with {\bf (a)} $g=1$ and {\bf (b)} $g=20$, and for the stronger quench of $\epsilon=5$ with {\bf (c)} $g=1$ and {\bf (d)} $g=20$. Each lower panel shows a magnified version of the evolution.}
\label{mossy0}
\end{center}
\end{figure}

To evaluate the finite-time thermodynamics of the system following the quench, we show in Fig.~\ref{mossy0} the LE for the same parameters as used in Fig.~\ref{mossy1}. The periodic nature of the echo is visible for the small quench ($\epsilon=2$) exhibiting breathing dynamics which are a consequence of the non-trivial energy shifts the system caused by the interaction~\cite{alon,schmelcher,gudyon}. In this case the frequency of the fast oscillations are given by the energy of the dominant quasi-particle peak in the corresponding spectral functions, which are governed by the ground state energy of the quenched state. The slower frequency envelope is a consequence of the finite interactions which cause a splitting of the first excited state into a pair of COM and REL states of comparable energy (see inset of Fig.~\ref{spec} {\bf (a)} and {\bf (b)}). The interference of these two states causes the breathing observed in the LE and the difference in their energy controls the breathing frequency. For the larger quench ($\epsilon=5$) it is clear that this beating is destroyed resulting in orthogonality and further destructive interference effects from the contributions of higher energy states.

\begin{figure}[t]
\begin{center}
{\bf (a)}\hskip0.45\columnwidth{\bf (b)}
\includegraphics[width=0.45\columnwidth]{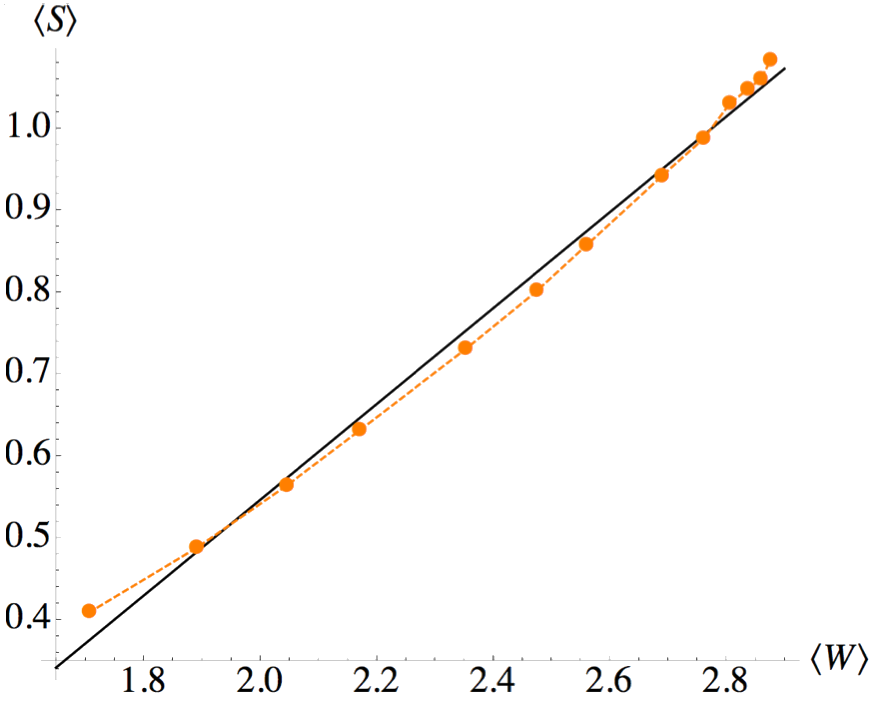}~~\includegraphics[width=0.45\columnwidth]{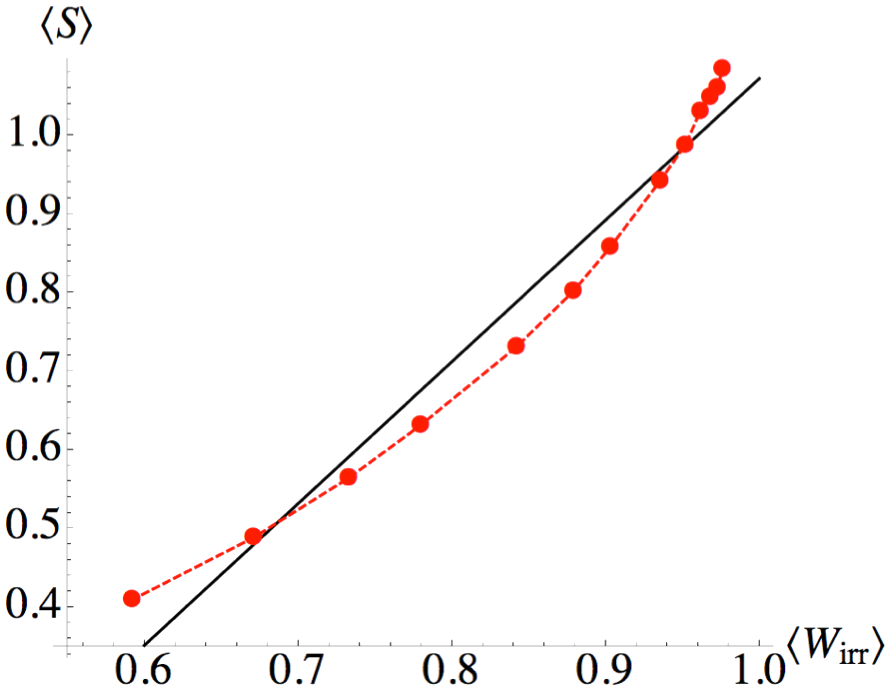}\\
{\bf (c)}\hskip0.45\columnwidth{\bf (d)}
\includegraphics[width=0.45\columnwidth]{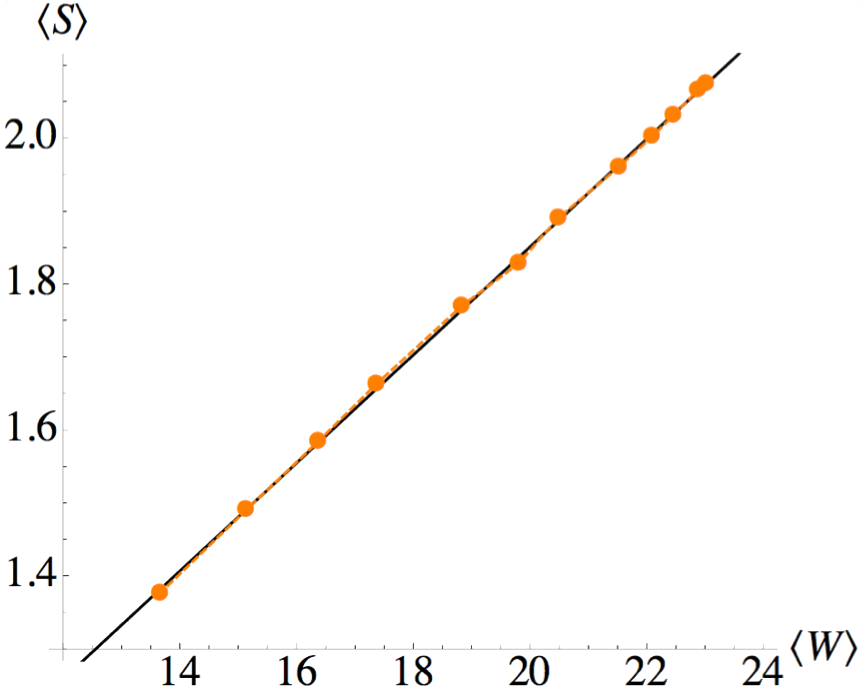}~~\includegraphics[width=0.45\columnwidth]{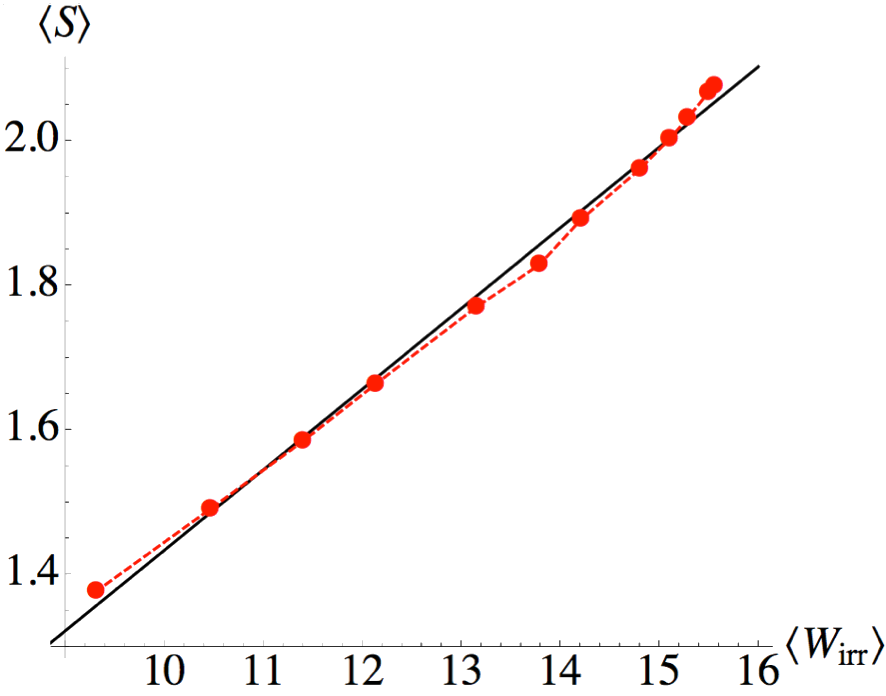}\\
\caption{Time-averaged vNE vs. (irreversible) work done on the system of two trapped bosonic atoms subjected to a quench of the trapping frequency. In all panels each point represents the value taken by the pair at a set value of $g$. The (black) straight lines show the result of a linear fit of the data points. Panel {\bf (a)} shows the behaviour of $\langle S\rangle$ against $\langle W\rangle$ and panel {\bf (b)} shows the behaviour of $\langle S\rangle$ against $\langle W_{\rm irr}\rangle$ for $\epsilon=2$. Panels {\bf (c)} and {\bf (d)} are for $\epsilon=5$.}
\label{mossy3}
\end{center}
\end{figure}

In the remainder of this Section we concentrate on the behaviour of the irreversible work $\langle W_{\rm irr}\rangle$ and its connection with the amount of entanglement created between the two atoms and the amount of work dissipated. 
To this aim, we calculate the mean vNE
\begin{equation}
\langle S\rangle=\frac{1}{\tau}\int^\tau_0S(t)dt,
\end{equation}
which is time-averaged over an interval $\tau$ that is long enough to include many periods of oscillation. In Fig.~\ref{mossy3} we show a clear qualitative link between $\langle S\rangle$ and both $\langle W \rangle$ and $\langle W_{\rm irr}\rangle$. We find all quantities follow the same qualitative behaviour (predominately linear, although exhibiting a slight systematic curvature) as a function of the interaction strength, thus suggesting a causal link between these figures of merit. This is intriguing as it evokes a possible cost function-role played by the thermodynamic irreversibility in the establishment of quantum correlations within an interacting system.

It is worth noting that a number of recent works have managed to establish a rigorous link between the appearance of correlations (both quantum and classical) and the associated thermodynamic cost~\cite{huber} (albeit applying a separate formalism to the one considered here), that is complementary to our analysis. Indeed, our results suggest a significant role for thermodynamic work in the establishment of quantum correlations between the atoms. Such a connection has previously been highlighted in the context of spring-like coupled bosons, but it was limited to Gaussian states and quadratic evolutions~\cite{Carlisle}. Our results go significantly beyond these restrictions as they address the case of contact-like bosonic interactions, which are highly experimentally relevant. In the following we will use this study on two coupled bosons as a benchmark for contact-like couplings among multiple atoms.

\section{Three trapped atoms}
\label{three}
We now extend the Hamiltonian model addressed so far to the more complex case of a one-dimensional mixture of two identical bosons of the same species X, whose coordinates will be indicated as $x_1$ and $x_2$, and one impurity atom of a different species Y, with coordinate $y$. We assume that all atoms have the same mass $m$ and are trapped with the same oscillator frequency $\omega$. The interactions are  of contact form and characterised by the intra- and inter-species coupling constants $g_{\mathrm{X}}$ and $g_{\mathrm{XY}}$. In this situation, the Hamiltonian reads  
\begin{equation}
\label{eq:Hamiltonian}
{\cal H}=-\frac12 \left[\sum^2_{j=1}\left(\partial^2_{x_j^2}{-}x_j^2\right){+}\Big(\partial^2_{y^2}{-}y^2\Big)\right] +g_\mathrm{X} \delta (x_1-x_2)+ g_\mathrm{XY}  \sum^2_{j=1}\delta (x_j{-}y),
\end{equation}
where the dimensionless coordinates are defined by rescaling energies and coupling rates as above.
Note that the eigenfunctions of Eq.~(\ref{eq:Hamiltonian}) have to  be symmetric with respect to 
the exchange of the X bosons, but no symmetry restriction for the interchange of the X atoms with the Y atom exists. A detailed study of the forms and properties of the eigenstates $\Psi(x_1,x_2,y)$  of Eq.~(\ref{eq:Hamiltonian}), focusing on the degeneracies of the spectrum, is given in the supplementary material. 

The RDM for the single atoms can be calculated by taking the partial trace of the state of the three-atom system over two of the atoms and is given for one of the X atoms by
\begin{equation}
\rho^X(x,x')=\int d\,x_2\,d\,y\,\psi(x,x_2,y)\psi(x',x_2,y) =\! \sum _k f_k^{\mathrm{X}}(x')f_k^{\mathrm{X}}(x)\lambda_k^{\mathrm{X}}.
\end{equation}
Analogously, for the impurity atom of species Y, we have
\begin{equation}
\rho^Y(y,y')=\int d\,x_1\,d\,x_2\,\psi(x_1,x_2,y)\psi(x_2,x_2,y') =\! \sum _k f_k^{\mathrm{Y}}(y')f_k^{\mathrm{Y}}(y)\lambda_k^{\mathrm{Y}}.
\end{equation}
Here the functions $f_k^{\mathrm{X,Y}}$ are the natural orbitals that diagonalise the RDMs with natural orbital occupations $\lambda_k^{\mathrm{X,Y}}$. The vNE of the impurity or of one of the atoms of species X can then be calculated as
\begin{equation}
 S^{\mathrm{X,Y}}=
 -\sum_k\lambda_k^{\mathrm{X,Y}}\ln\lambda_k^{\mathrm{X,Y}}. 
\end{equation} 

\begin{figure}[t!]
\begin{center}
\includegraphics[width=0.48\columnwidth]{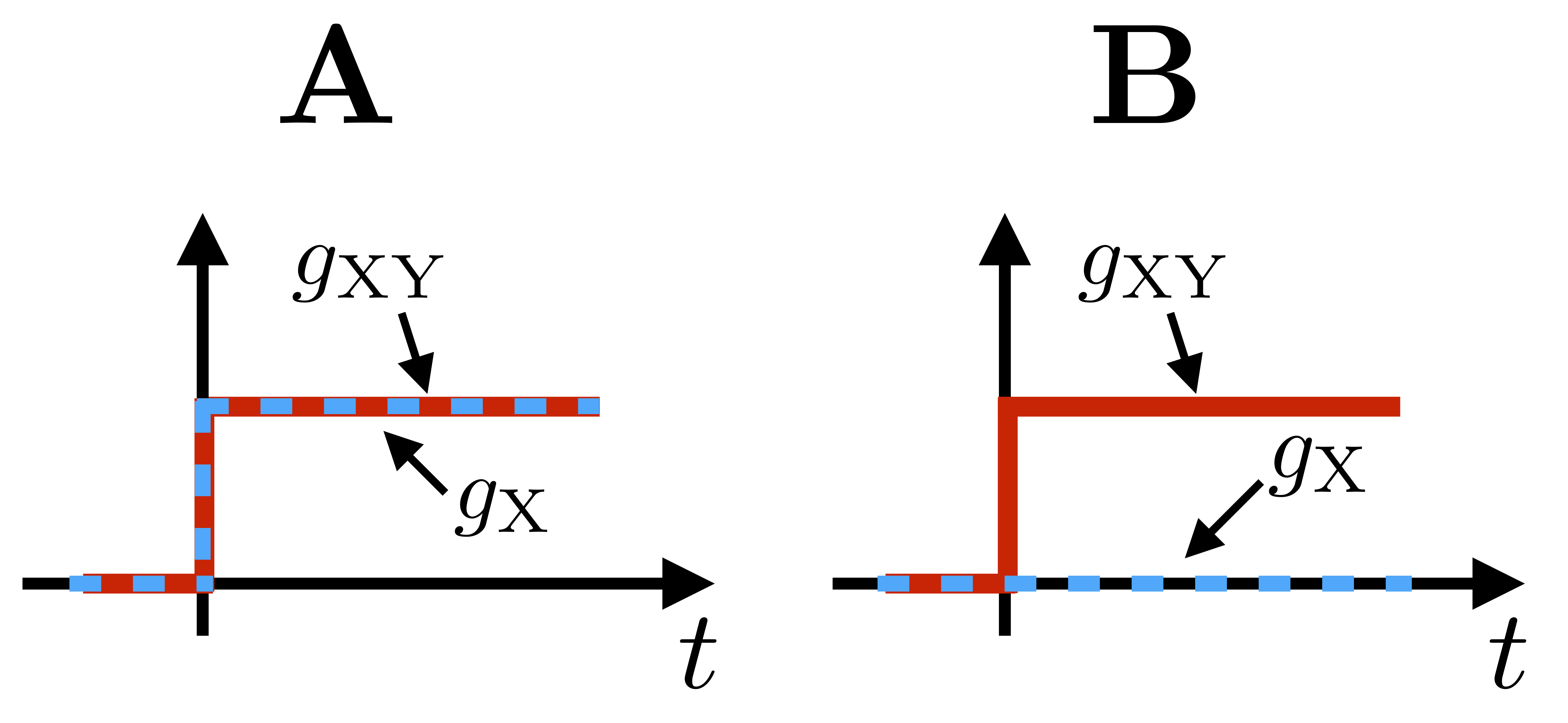}
\caption{Sketch of the two different quenching strategies, {\bf A} and {\bf B}, considered for the three-atom system.}
\label{figQ}
\end{center}
\end{figure}

As mentioned previously, for any finite interaction strength, a quench in the trapping frequency results in an effective change of the interaction strength. While for the trapped molecule a quench of either parameter led to the same behaviour, the presence of two coupling constants in the current system complicates the matter. In order to clearly identify the role atomic interactions play in the dynamics, we therefore restrict ourselves in the following to quenching the interaction strengths directly. It is worth noting that, when $g_{\mathrm{X}}=g_{\mathrm{XY}}=0$ or $g_{\mathrm{X}}=g_{\mathrm{XY}}\approx \infty$\footnote{In our simulations $g=20$ is large enough to effectively reach the TG limit.}, the behavior of the three-atom system is qualitatively the same as the single and two atom cases, respectively.

For our two-component system there are in principle five different strategies for quenching the coupling constants $(g_{\mathrm{X}},g_{\mathrm{XY}})$, but in the following we will focus on just two strategies as illustrated in Fig.~\ref{figQ}, as these encompass the most relevant features of the thermodynamic properties of the system (see Appendix A for a discussions of the other strategies)

\vspace{5pt}
\begin{tabular}{ l l l }
  \bf{A}: & $(0,0)\rightarrow (g,g)$,\\
  \bf{B}: & $(0,0)\rightarrow (0,g)$.\\
\end{tabular}
\vspace{5pt}

\begin{figure}
\begin{center}
\includegraphics[width=0.98\columnwidth]{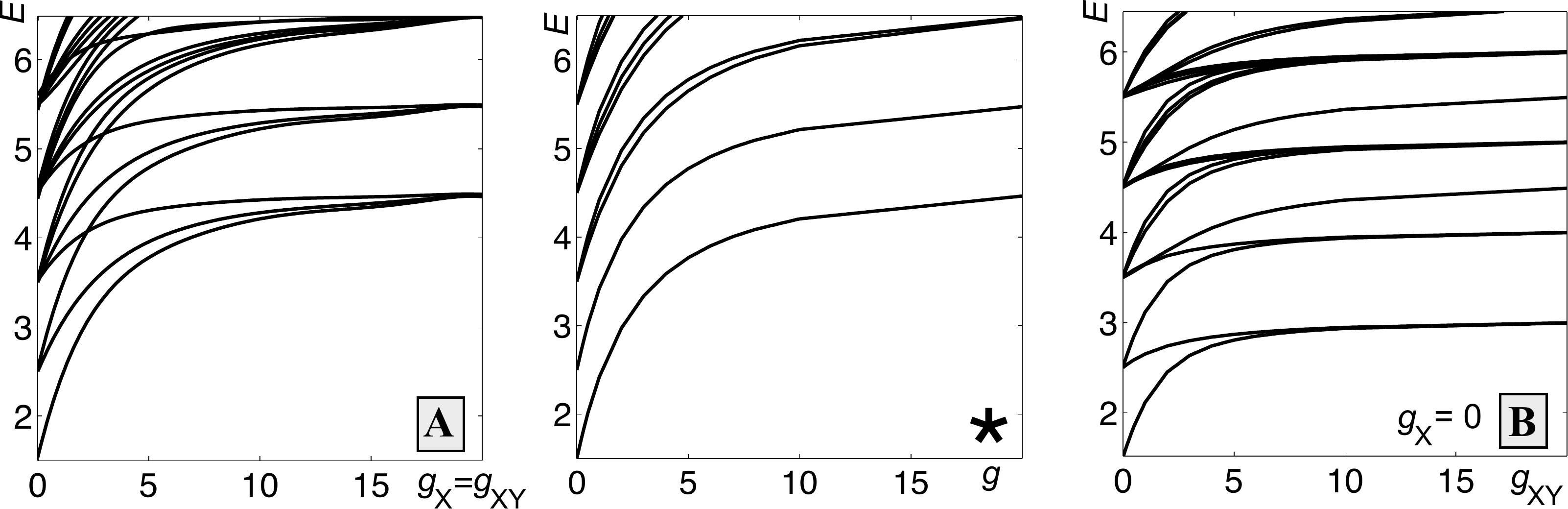}
\caption{Energy eigenspectra for the quenches {\bf A}, ($*$) which denotes a quench of the interaction between three indistinguishable bosons, and quench {\bf B}. The situation {\bf A}, where symmetric scattering between two of the atoms is required (and absent for the scattering with the third atom) should be compared to ($*$), where the existence of a symmetry condition between all atoms was assumed. In situation {\bf B} all symmetry requirements are absent and therefore exhibits a unique energy spectrum.}
\label{fig1} 
\end{center}
\end{figure}

\noindent
The energy spectra corresponding to the adiabatic version of these quenches are shown in Fig.~\ref{fig1} \cite{Garcia-March:14b,Dehkharghani:15, Garcia-March:16,ZinnerEPL} and the visible differences are directly related to the distinct outcomes of the particular quenching protocols. The spectrum for situations adhering to strategy {\bf A} shows the emergence of three-fold degenerate states in the limit of infinite interaction strengths, while in the same limit strategy {\bf B} leads to a two-fold degeneracy. The differences among these energy spectra make it clear that each quenching protocol provides distinct thermodynamic and density evolutions, which we will discuss below. We will show that both the entanglement and the thermodynamical quantities of interest are strongly affected by the chosen quenching strategy and we highlight a link between these quantities and the behaviour of the atomic density profiles. We remark that a consistent feature of all results is the periodic nature of these functions, which is due to the gas refocussing in its external harmonic oscillator potential at approximate multiples of the trap frequency.

\subsection{Thermodynamic quantities}
\label{thermothree}
\begin{figure}[t!]
\begin{center}
\includegraphics[width=0.85\columnwidth]{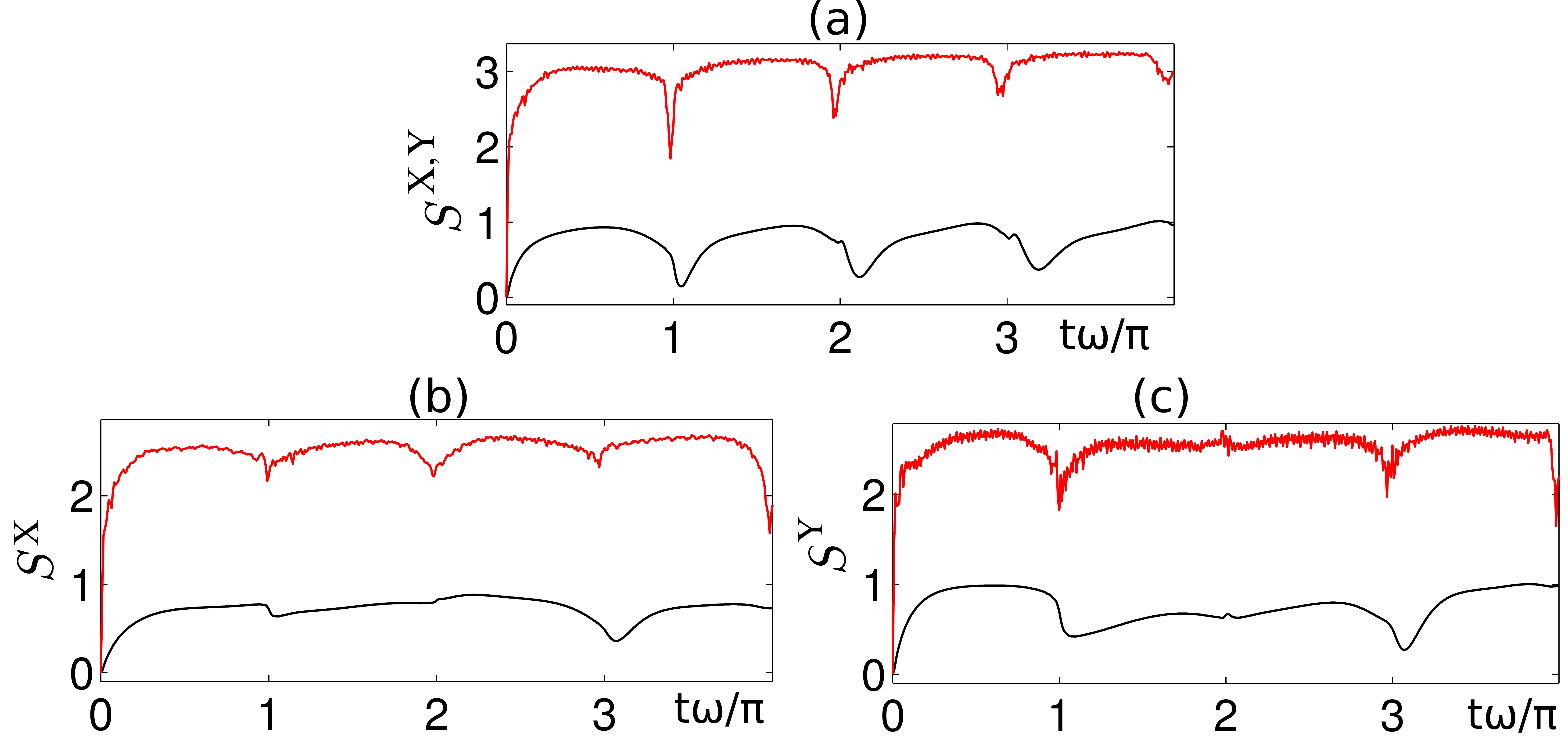}
\caption{{\bf (a)} {Evolution of the vNE after a quench of type {\bf A}, where $g_{\mathrm{X}}$ and $g_{\mathrm{XY}}$ are quenched from 0 to 2 or to 20 (black  and red curve, respectively). }   This quantity is equal for  either species. {\bf (b)}  and {\bf (c)}  Entropies $S^{\rm X}$ and $S^{\rm Y}$ for $g_{\mathrm{X}}=0$ with $g_{\mathrm{XY}}$ quenched to $2$ or $20$ (black and red curves respectively) following the quenching strategy {\bf B}.}
\label{fig3}
\end{center}
\end{figure}

For strategy {\bf A} the behaviour of the vNE is shown in Fig.~\ref{fig3} {\bf (a)}. 
It is worth noting that the behaviour for this quantity is the same whether two X atoms or one X and one Y atom are traced out. This  can be understood by realising that when quenching the two coupling constants to the same value, the system behaves as one composed of three indistinguishable atoms. It indicates that only the energy levels that both systems have in common are affected by the quench and there is no difference between the respective RDMs. A more formal argument using group theory is presented in Appendix B. We can also see that the entanglement increases with the amplitude of the quench and that dips appear with a periodicity close to multiples of the trapping frequency. These are more harmonic and narrower when the system is quenched close to the TG regime ($g=20$), as it can then be mapped onto non-interacting fermions. In Figs.~\ref{fig3} {\bf (b)} and {\bf (c)} we show the evolution of the vNEs for strategy {\bf B}, which now differ for the atoms of different species. Again, the vNE grows for stronger quenches, but the periodic dips are less pronounced compared to the ones observed with strategy {\bf A}. This is not surprising, as the spectrum for this situation is denser and therefore more states contribute to the evolution, which makes perfect refocussing less likely.

\begin{figure}
\begin{center}
\includegraphics[width=0.9\columnwidth]{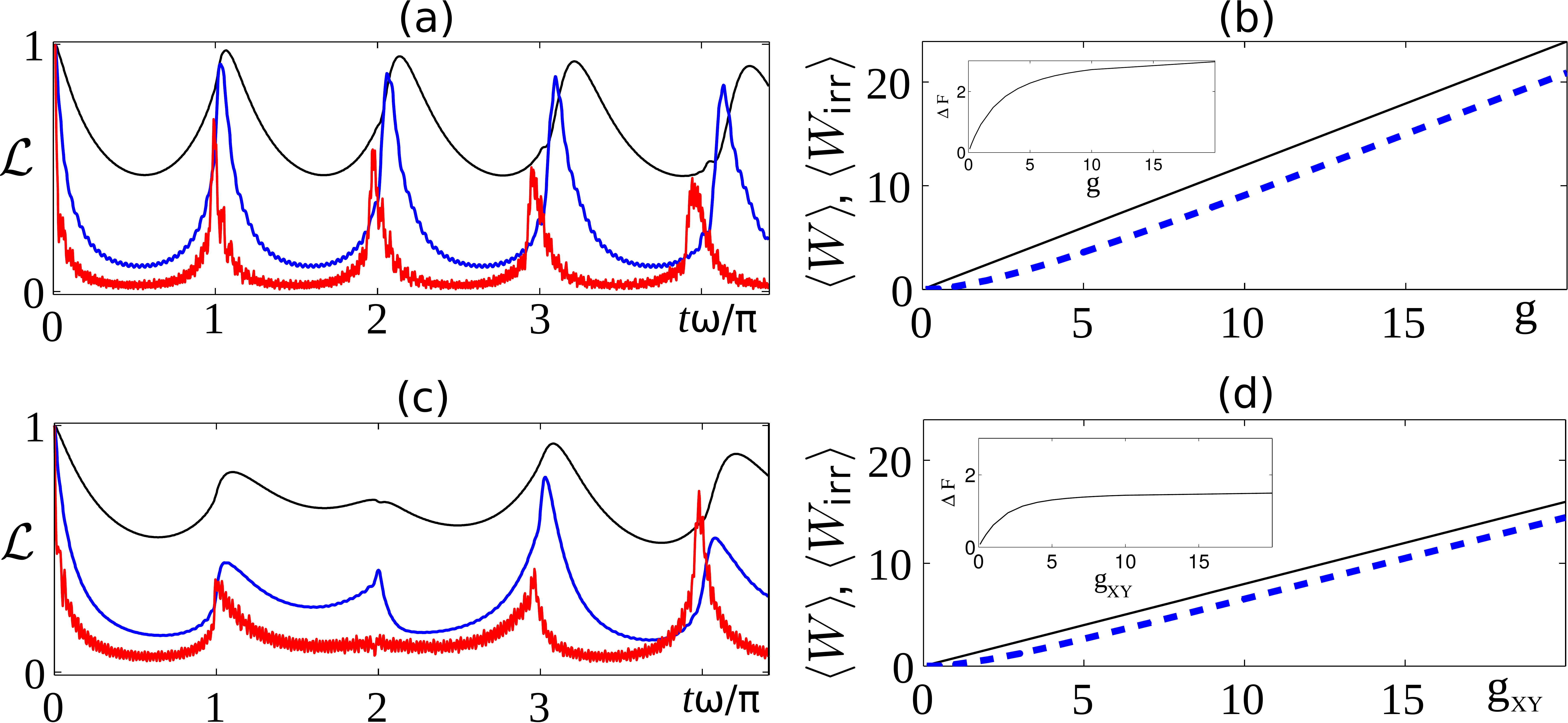}
\caption{Temporal behavior of the LE for the different quenching strategies. {\bf (a)} Quench {\bf A} with $g_{\mathrm{X}}=g_{\mathrm{XY}}=$ 2 (black), 6 (blue), 20 (red); {\bf (c)} quench {\bf B} with $g_{\mathrm{XY}}=$ 2 (black), 6 (blue), 20 (red) and $g_{\mathrm{X}}=0$; Panels {\bf (b)} and {\bf (d)} show the corresponding average work (solid black lines) and irreversible work (dashed blue lines) for such quenches, with the insets showing the free energy change $\Delta F$ against the quenching amplitude.} 
\label{fig8}
\end{center}
\end{figure}

The LEs following the quenches are shown in Figs.~\ref{fig8} {\bf (a)} and {\bf (c)} and, similar to the behaviour of the vNE above, they display regular revivals, whose period moves closer to the harmonic oscillator time scale of multiples of $\omega t/\pi$ for large quench amplitudes (when the system is quenched to the TG regime). The different behaviours for each quench are most clearly visible by looking at the behaviour of the LE around the revivals. Let us first discuss quenches into the TG regime, for which the LE always displays a periodicity.  For strategy {\bf A} the revivals have a period $\omega t/\pi$, which is due to the evolution being governed by the energy differences between the quenched states and the initial state, $\Delta E = E'_n-E_0 = q$, where $q$ is an integer as the trapping frequencies do not change. For case  {\bf B}, the LE is more varied and shows periodicities with $4 \omega t/\pi$ and $2 \omega t/\pi$. This is a consequence of the symmetry considerations discussed in the supplementary material, which results in energy differences (a combination of integer and half integer $\Delta E$) for strategy {\bf B}. For weaker quenches, $g<20$, the LEs show complex temporal patterns which is a sign that the energy spectrum is not as degenerate as in the TG limit.

The correspondence between the LE and the characteristic function of the work distribution allows us to evaluate the average work done on the system as a result of the quench (see  Figs.~\ref{fig8} {\bf (b)} and {\bf (d)}). Quite interestingly, we find that $\langle W\rangle$ depends linearly on the quenching strength in both settings, while the irreversible work $\langle W_{\rm irr}\rangle$ produced in such non-quasistatic processes behaves linearly only at larger values of the quench amplitude. This arises from the impossibility of the system to generate increasing amounts of `useful' work as the strength of the couplings grows. Indeed,  the free energy difference $\Delta F=\Delta E=E'_0-E_0$ (see insets of Figs.~\ref{fig8} {\bf (b)} and {\bf (d)}), levels off at large values of the interaction strengths, showing that the system soon saturates its capabilities to produce work that can be usefully extracted. Large quenching amplitudes are thus associated with an increasing degree of irreversibility. These considerations allow us to identify an optimal configuration of quenching, associated with moderate quenching amplitudes, that without requiring sophisticated control techniques is able to attain the maximum allowed useful work without generating an unbounded amount of thermodynamic irreversibility.

\begin{figure}[t!]
\begin{center}
\includegraphics[width=0.9\columnwidth]{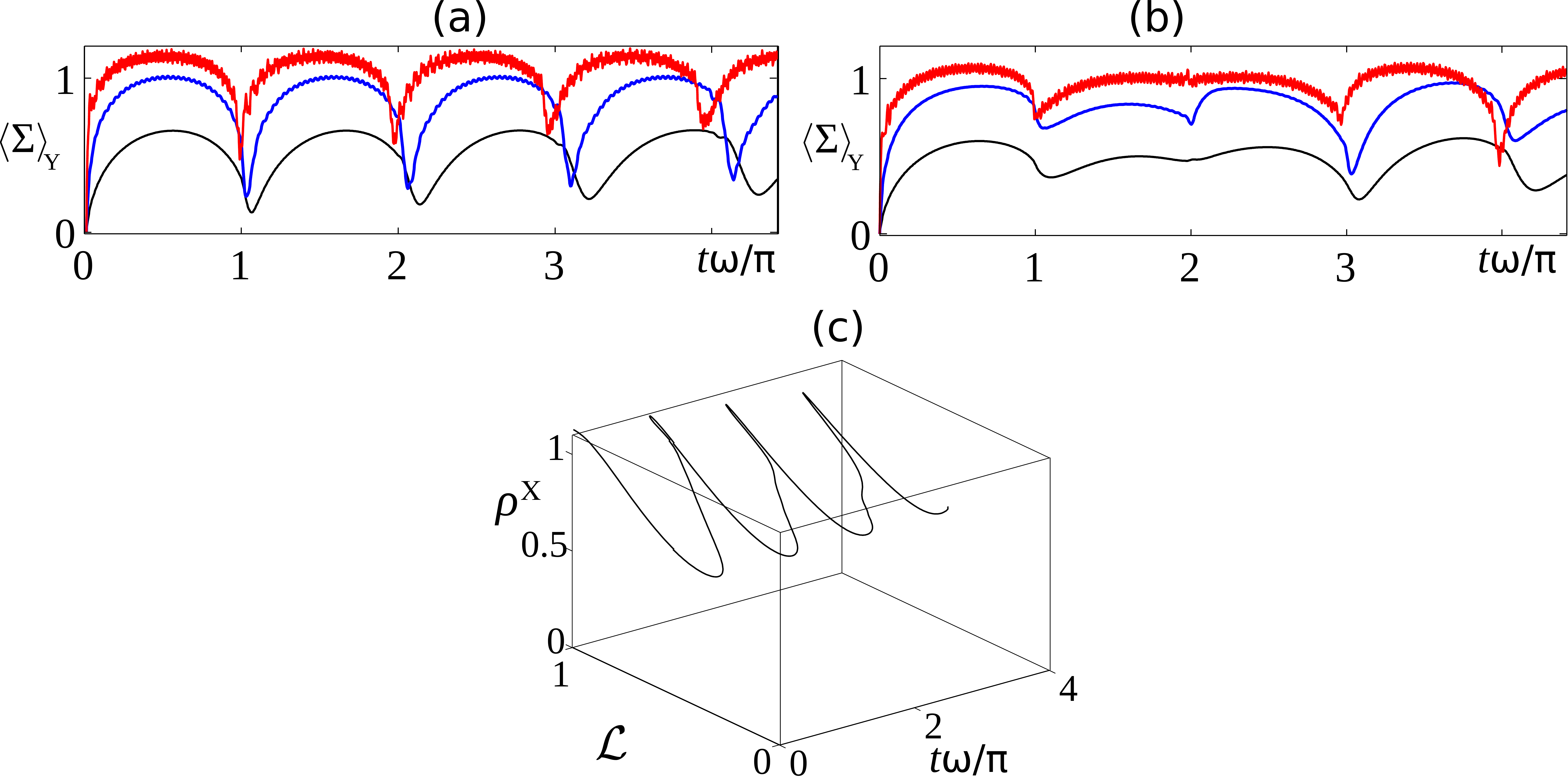}
\caption{Lower bounds of the dynamical irreversible entropies produced with the different quenching strategies. {\bf (a)} Quench {\bf A} with $g_{\mathrm{X}}=g_{\mathrm{XY}}=$ 2 (black), 6 (blue), 20 (red); {\bf (b)} Quench {\bf B} with $g_{\mathrm{XY}}=$ 2 (black), 6 (blue), 20 (red) and $g_{\mathrm{X}}=0$; In panel {\bf (c)} the density at the centre of the trap $x=x'=0$ of one of the atoms of species X can be seen as a function of the dimensionless  time and the LE for quenching strategy {\bf A}, with $g=2$. } 
\label{fig9}
\end{center}
\end{figure}

The lower bound of the average irreversible entropy produced as a result of the quenches is shown in Fig.~\ref{fig9} {\bf (a)}-{\bf(b)} and it is apparent that $\langle\Sigma\rangle_B$ closely follows the same temporal trend as ${\cal L}$. This is not surprising, given that for a time-independent Hamiltonian and a system prepared in the ground state of the initial Hamiltonian, the LE coincides with the state fidelity, and the latter directly enters the definition of the Bures angle. {In Fig.~\ref{fig9} {\bf (c)} we show a qualitative link between the behaviour of the LE and the single atom density of the system. The visible regular dependence suggests the possibility that the behaviour of thermodynamically relevant quantities, such as the irreversible entropy production, can be inferred from experimentally accessible figures of merit such as the density profile, which we will explore in the next Section}.

\subsection{Density evolutions} 
 \begin{figure}[t]
\begin{center}
{\bf (a)}\hskip0.35\columnwidth{\bf (b)}
\includegraphics[width=0.99\columnwidth]{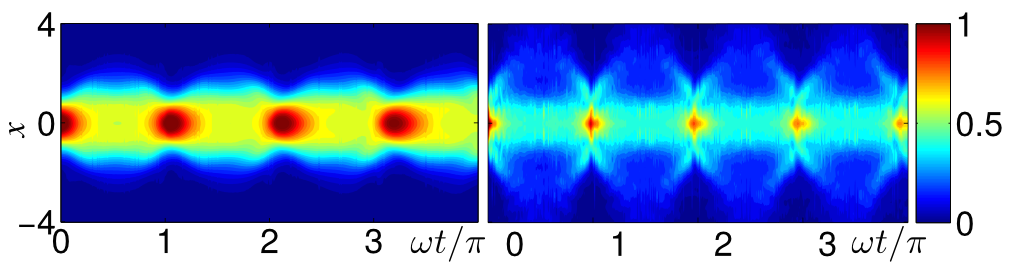}
\caption{Evolution of the density profile $\rho^X(x)$ for a quench of strategy {\bf A} for {\bf (a)} $g=2$ and {\bf(b)} $g=20$.}
\label{fig2bis}
\end{center}
\end{figure}

\begin{figure}[t!]
\begin{center}
{\bf (a)}\hskip0.5\columnwidth{\bf (c)}
\includegraphics[width=.98\columnwidth]{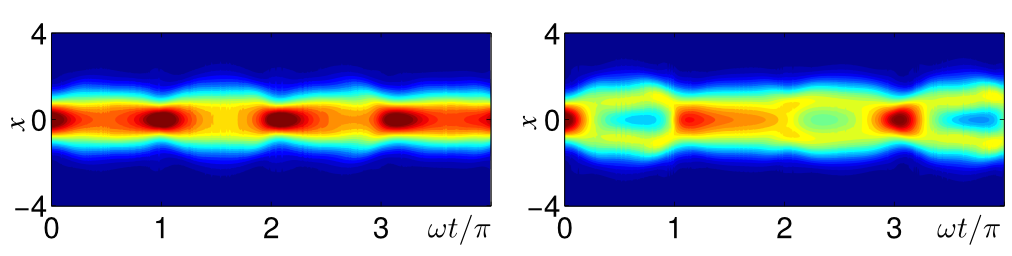} \\
{\bf (b)}\hskip0.5\columnwidth{\bf (d)}
\includegraphics[width=.98\columnwidth]{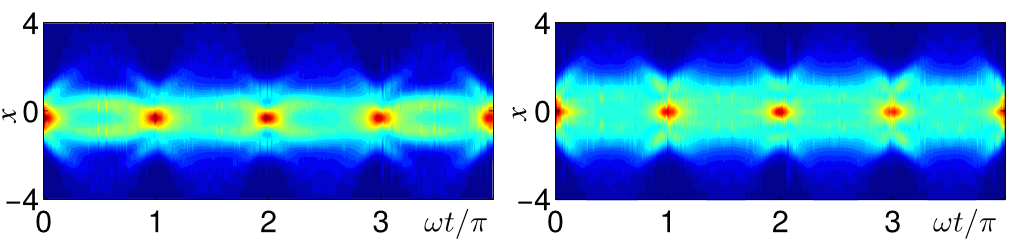}
\caption{Density evolution of the two X atoms [{\bf (a)} and {\bf (b)}] and the impurity Y [{\bf (c)} and {\bf (d)}] for quenching strategy {\bf B} with $g_{\rm X}=0$ and $g_{\mathrm{XY}}$ quenched to $g=2$ and $g=20$ [top and bottom row of plots, respectively]. Same color-scale as in Fig.~\ref{fig2bis}.} 
\label{fig4}
\end{center}
\end{figure}

The evolution of the density profile after quench {\bf A} for two different strengths is shown in Fig.~\ref{fig2bis} and we note that, as expected, the profiles obtained from the RDM for X and Y are identical. We can see that the system  is localised around $x=0$ and oscillates at a frequency which depends on the strength of the quench. In fact, these oscillations mirror the appearance of the dips in the vNE and the peaks in the LE (see Figs.~\ref{fig3} {\bf (a)} and \ref{fig8} {\bf (a)}). A larger amplitude of the quench leads to narrower revival peaks, which again corresponds to tighter dips (peaks) in the vNE (LE).
 
For the situation of strategy {\bf B}, where $g_{\rm X}=0$ and $g_\mathrm{XY}$ is quenched, it was shown above that the LE and the vNE show complex dynamics for small quench values. As the two species are distinguishable in this case, we show in Fig.~\ref{fig4} the density evolutions for an atom of species X and Y separately. For small $g_\mathrm{XY}$, the density profile for species X remains localised in the center of the trap, while the impurity can {\it spread} to the edges of the distribution for the X atoms, forming a double peaked structure at certain times ~\cite{Garcia-March:14b}. The intrincate structure of the LE in this case can be related to  the irregular temporal evolution of the Y atom compared to the more periodic oscillations of the X atoms. However for larger $g_\mathrm{XY}$, the periodicity of the evolution becomes more regular for both species as the energy structure becomes more degenerate, resulting in complementary trends visible in the corresponding vNE and LE (see Figs.~\ref{fig3} {\bf(c)} and {\bf(d)} and \ref{fig8} {\bf(c)}).  

A connection between the dynamics of the density profile and the evolution of thermodynamically relevant quantities would allow insight into the finite-time thermodynamics of non-equilibrium processes without requiring measurements of hard-to-access variables~\cite{campisi}. In particular, Fig.~\ref{fig9} {\bf (c)} highlights the possibility of post-processing data acquired on the density profile  at the centre of the trap within one period of the evolution to infer the corresponding value taken by the LE.

\section{Conclusions}
\label{conclusion}

We have studied the finite-time thermodynamics of a small-sized gas of interacting bosonic atoms subjected to a sudden quench of the Hamiltonian parameters. By first reviewing the out-of-equilibrium dynamics of a quenched single atom we have confirmed that larger quenches lead to larger amounts of entropy produced which implies an increase in the amount of work and irreversible work injected into the system. However, the system only appears to evolve into an orthogonal state when the quench is infinitely large. For the two-atom case we have investigated the interesting role that particle interactions play. In particular, starting from the analytically tractable Tonks-Girardeu regime, we noted that for such infinitely repulsive bosons, the strong interaction enhances the entropy production and the system can now exhibit full orthogonality. We also established the extensive nature of work in this system. For finite interactions between the atoms, we have shown a clear qualitative link between the amount of (irreversible) work performed on the system and  the increase in the degree of inter-atomic entanglement. Moving into the multipartite case, and despite the significant increase in the complexity of the problem (as evidenced by the range of inherently different dynamics and sprectra observable simply by altering the initial interaction strengths), we have highlighted that the qualitative features of the two particle case appear to persist, i.e. the initial interactions strongly dictate the dynamical features. Finally, we have shown that the behaviour of the atomic density profile of a single atom can be a useful tool in exploring the non-equilibrium properties of a system, even in the case of complex multipartite systems.

\section*{Acknowledgements}
We thank Rosaria Lena, Gabriele De Chiara, and G. Massimo Palma for discussions related to the non-equilibrium thermodynamics of ultracold atomic systems. MAGM, and SC are grateful to the Okinawa Institute of Science and Technology for hospitality during the development of this project. This work was supported by the Okinawa Institute of Science and Technology Graduate University, the EU FP7 grant TherMiQ (Grant Agreement 618074), the John Templeton Foundation (Grant No. 43467), the UK EPSRC (EP/M003019/1). Part of this work was supported by the COST Action MP1209 ``Thermodynamics in the quantum regime". TF acknowledges support from the German Research Foundation (DFG, DACH project Quantum crystals of matter and light) and  BMBF (Qu.com). M.A.G.-M. acknowledges support from EU grants OSYRIS (ERC-2013-AdG Grant No. 339106), SIQS (FP7-ICT-2011-9 No. 600645), EU STREP QUIC (H2020-FETPROACT-2014 No. 641122), EQuaM (FP7/2007-2013 Grant No. 323714), Spanish Ministry grant FOQUS (FIS2013-46768-P),  the Generalitat de Catalunya project 2014 SGR 874, and Fundaci\'o Cellex. Financial support from the Spanish Ministry of Economy and Competitiveness, through the ``Severo Ochoa" Programme for Centres of Excellence in R\&D (SEV-2015-0522) is acknowledged.

\section*{Appendix A: Analysis of other Quenching Strategies}  
\label{appC}

\begin{figure}
\begin{center}
\subfigure[]{\includegraphics[width=0.68\columnwidth]{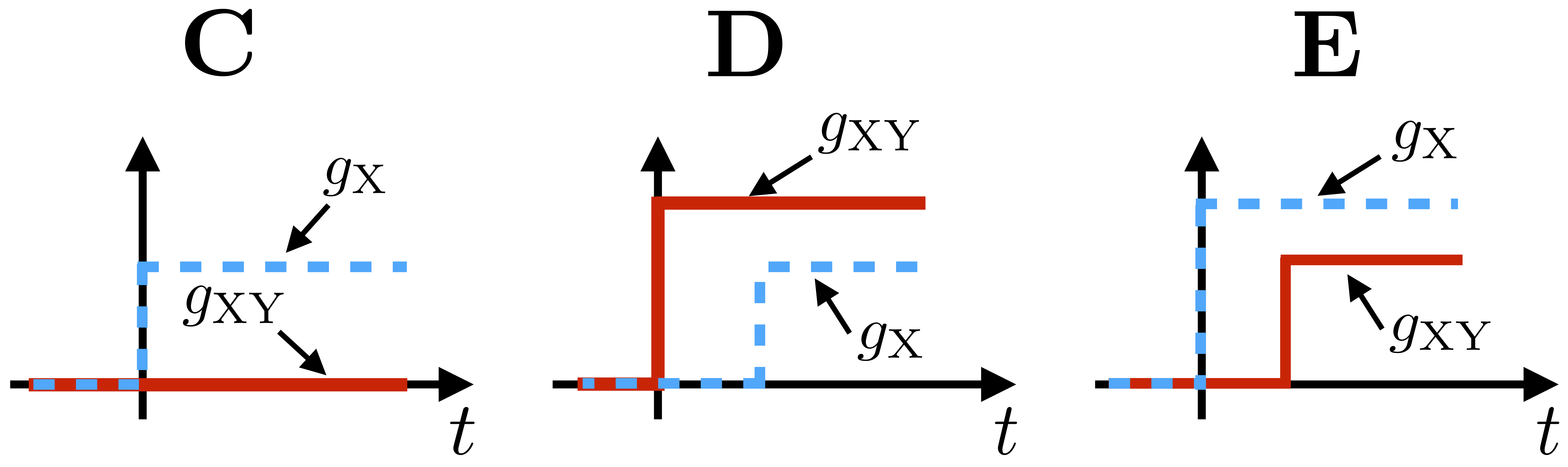}}
\subfigure[]{\includegraphics[width=0.98\columnwidth]{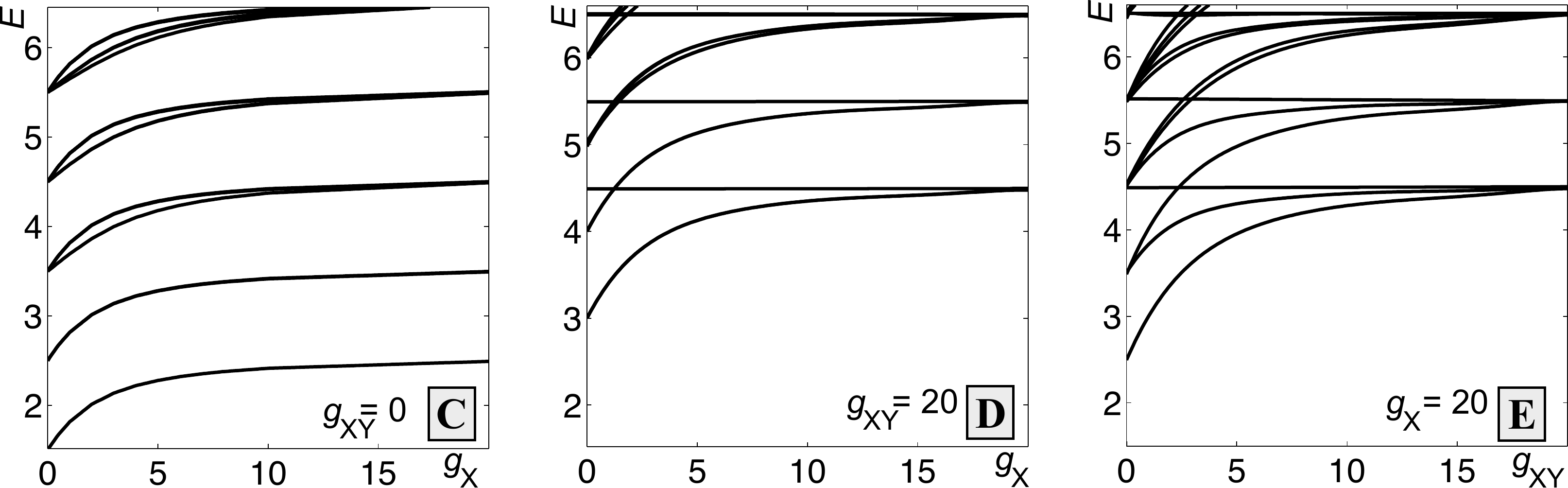}}
\caption{(a) Sketch of the three other quenching strategies considered for the three-atom system. The quenches {\bf D} and {\bf E} could in principle lead to the same final values for $g_{\rm X}$ and $g_{\rm XY}$. (b) Corresponding energy eigenspectra for these systems with different scattering symmetry requirements. The sprectra relating to quench {\bf E} fixes the intra-species coupling strength, $g_{\rm X}$, and changes the inter-species coupling strength, $g_{\rm XY}$, while for the spectra {\bf C} and {\bf D} the opposite is the case. In situation {\bf C} only scattering with a symmetry requirement is present and it should therefore be compared with situation {\bf B} where all symmetry requirements are absent. Finally, in situation {\bf D} the symmetric scattering is fixed and the scattering without the symmetry condition is varied, which should be compared to the situation {\bf E} where the opposite is the case.}
\label{figSPEC} 
\end{center}
\end{figure}

While in the main body we focused on two quenching strategies for the 3-atom system, in this appendix we briefly address the remaining protocols for quenching the coupling constants $(g_{\mathrm{X}},g_{\mathrm{XY}})$ which are shown in Fig.\ref{figSPEC}:

\vspace{5pt}
\begin{tabular}{ l l l }
  \bf{C}: & $(0,0)\rightarrow (g,0)$,\\
  \bf{D}: & $(0,\infty)\rightarrow (g,\infty)$,\\
  \bf{E}: & $(\infty,0)\rightarrow (\infty,g)$.\\
\end{tabular}
\vspace{5pt}

\noindent
In strategy {\bf C} the interaction between the two species remains zero. Thus the impurity atom Y acts as a spectator only, and qualitatively this scenario is identical to the analysis presented in Sec.~\ref{molecule}. Furthermore, strategy {\bf E} corresponds to a TG molecule suddenly interacting with an impurity atom which has been extensively studied in Ref.~\cite{Campbell:14}. Therefore, we finally comment on strategy {\bf D}, which shares several features in common with strategy {\bf B}, with the important distinction that the atoms are initially interacting at $t<0$, and therefore many natural orbitals have non-zero occupation even before the quench~\cite{Garcia-March:14b,Girardeau:07, Deuretzbacher:14}.

\begin{figure}[t]
\begin{center}
\includegraphics[width=0.9\columnwidth]{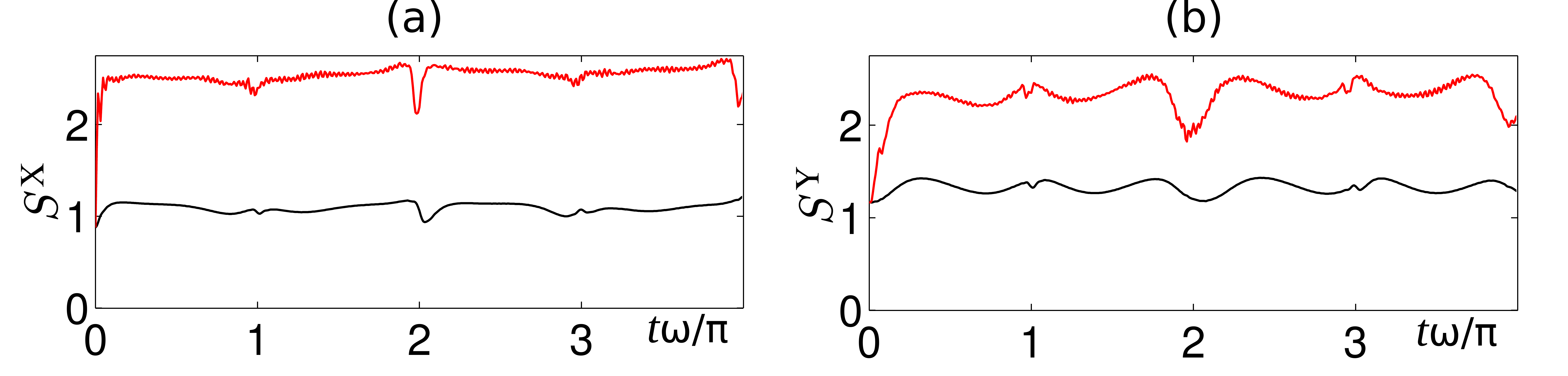}
\caption{{\bf (a)} and {\bf (b)} Entropies $S^{\rm X}$ and $S^{\rm Y}$ following the quenching strategy {\bf D}, where $g_{\mathrm{XY}}=20$ and $g_{\mathrm{X}}$ is quenched from 0 to $2$ or $20$ (black and red curves respectively).}
\label{fig3App}
\end{center}
\end{figure}

\begin{figure}[t]
\begin{center}
\includegraphics[width=0.9\columnwidth]{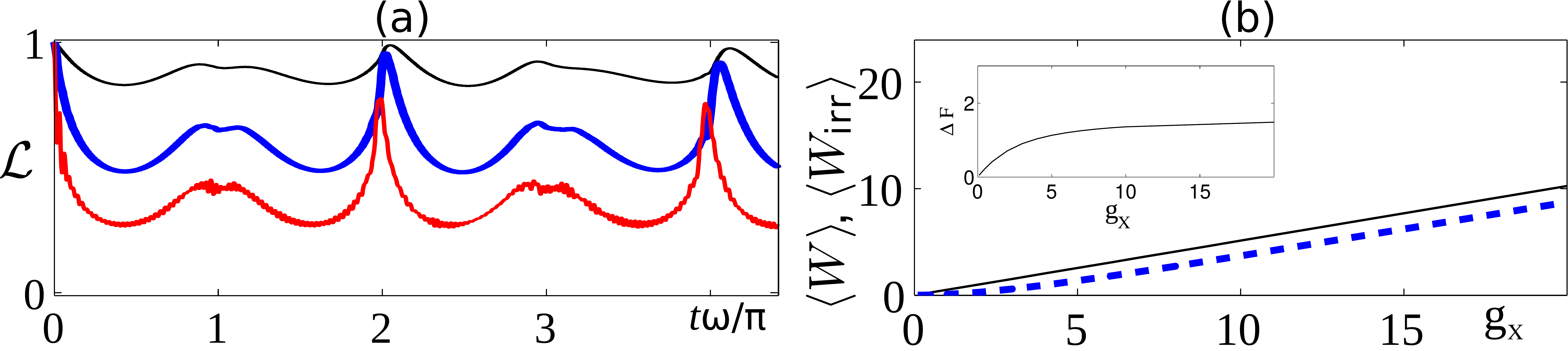}
\caption{{\bf (a)} Temporal behavior of LE for Quench {\bf D} with $g_{\mathrm{X}}=$ 2 (black), 6 (blue), 20 (red) and $g_{\mathrm{XY}}=20$; Panel {\bf (b)} shows the corresponding average work (solid black lines) and irreversible work (dashed blue lines) with the inset showing the free energy change $\Delta F$ against the quenching amplitude. We remark that the irreversible entropy behaves qualitatively similar to LE.} 
\label{fig8App}
\end{center}
\end{figure}

The dynamical behaviour of  the vNE for strategy {\bf D}  is shown in Figs.~\ref{fig3App} {\bf (a)} and {\bf (b)}. As the interaction between the two species is already large at the beginning, the initial values of the correlations are now finite and the quench increases them to a similar level as in the cases considered in the main text. Similarly, periodic dips appear again around the refocussing time of the harmonic trap. Figs.~\ref{fig8App} {\bf (a)}  shows the LE that, similar to strategy {\bf B}, exhibits periodicities around $4 \omega t/\pi$ and $2 \omega t/\pi$ due to the energy differences of $\Delta E=q+1/2$. We see from panel Fig.~\ref{fig8App} {\bf (b)} the behaviour of the average (irreversible) work and free energy are qualitatively the same as for strategy {\bf B}.

\begin{figure}[t]
\begin{tabular}{c}
{\bf (a)}\hskip0.5\columnwidth{\bf (c)}\\
\includegraphics[width=.93\columnwidth]{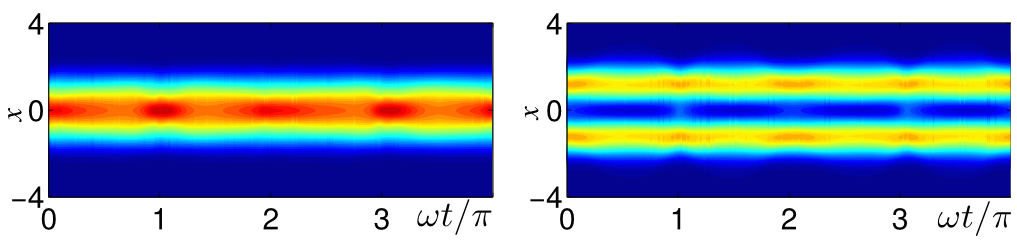} \\
{\bf (b)}\hskip0.5\columnwidth{\bf (d)}\\
\includegraphics[width=.93\columnwidth]{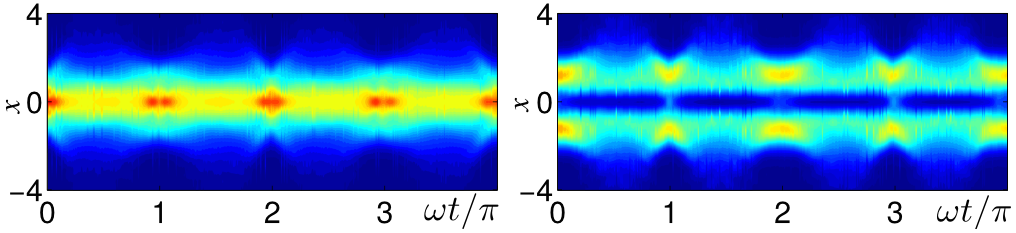}
\end{tabular}
\caption{Evolution of the density profile for an X atom [panels {\bf (a)} and {\bf (b)}] and the impurity Y [panels {\bf (c)} and {\bf (d)}] resulting from the implementation of strategy {\bf D}. Here $g_{\mathrm{XY}}=20$ is constant and we consider $g_{\mathrm{X}}$ quenched to $g=2$ and $g=20$   [top and bottom row of plots, respectively]. Same color-scale as in Fig.~\ref{fig2bis}.} \label{fig6App}
\end{figure}

Fig.~\ref{fig6App} shows the evolution of the density profile for a quench of $g_\mathrm{X}$ taken from $0$ to a value that is either much smaller than $g_{\rm XY}$, or comparable to it. The phenomenology is strikingly different for both cases. For $g_{\rm X}\ll g_{\rm XY}$  the density profile for the X atoms is peaked at center of the trap, while the impurity Y has a double-peak structure which is localised at the sides of the density of X. This distribution shows only a weak temporal change and the separation is maintained, which corresponds to the flat vNE of the individual species (see Fig.~\ref{fig3App} {\bf (a)} and {\bf (b)}). The case in which the final value of $g_{\rm X}$ is comparable to the inter-species coupling rate shows more pronounced temporal oscillations, which are also seen in the behaviour of the vNE . The atomic species are strongly correlated regardless of the strength of the quench. However, large quenching amplitudes result in dips of the vNE at the refocusing time of the density profile that are much less pronounced than those occurring at small final values of $g_{\rm A}$.

\section*{Appendix B: Comparison with indistinguishable atoms}  
\label{appB}
It is also interesting to compare the evolution of the density, the vNE, and occupation of the natural orbitals shown in Figs.~\ref{fig3} and \ref{fig2bis} with that of a system of three indistinguishable atoms. The energy spectrum for such a system is  shown in Fig.~\ref{fig1} {($*$)}, and clear differences from that of two atoms plus a third distinguishable one are visible. In the latter case triple degenerate states occur in the limit $g_\mathrm{X},g_\mathrm{XY}\to\infty$ (see Fig.~\ref{fig1} quench {\bf A}) and a discrete group theory analysis presented in Refs.~\cite{Garcia-March:14b,Harshman:12} showed that these three degenerate states belong to different irreducible representations of the group. The discrete group to which these solutions belong is the discrete rotational group of order 2, $\mathcal{C}^2$, restricted by the bosonic symmetry under interchange of the two indistinguishable atoms.  Indeed, for all finite values of the coupling constants, all  wave functions  can be classified according to the possible irreducible representations of this group. In Ref.~\cite{Garcia-March:14b} it was shown that the ground state for the three indistinguishable atoms  was the same as the 2+1 case for all values of $g=g_\mathrm{X}=g_\mathrm{XY}$. This is not too surprising, as these solutions obey all symmetries under interchange of two atoms required by the  three-indistinguishable atoms which coincide with the ones required by the corresponding irreducible representation of the group in the 2+1 system. 

By comparing the dynamical evolution of the  density, the vNE, and the occupations of the natural orbitals for a system of three-indistinguishable atoms with the 2+1 setting, we find that they all coincide. The reason for this is that the initial state is a non-interacting Gaussian state with certain symmetries, which corresponds to the absence of a change in the sign of the wave function when any pair of atoms interchanges their coordinates. Therefore it belongs to a definite irreducible representation of the  $\mathcal{C}^2$, restricted by the bosonic symmetry under the interchange of the two X atoms~\cite{Garcia-March:14b}. If the system has this symmetry initially, the dynamical evolution has to conserve it, so only part of the energy spectra in the 2+1 case plays a role in the evolution. This part of the energy spectra is exactly the same as in the case of three indistinguishable atoms.

\end{document}